\documentclass{aa}  

\usepackage{graphicx}
\usepackage{longtable}
\usepackage{caption}
\usepackage{txfonts}
\usepackage{amsmath,amssymb}
\begin{document}

   \title{Scaling laws of quasi-periodic pulsations in solar flares}


   \author{C. E. Pugh\inst{1}
          \and
          A.-M. Broomhall\inst{1}
          \and
          V. M. Nakariakov\inst{1,2}
          }

   \institute{Department of Physics, University of Warwick,
              Coventry, CV4 7AL, UK\\
              \email{a-m.broomhall@warwick.ac.uk}
         \and
             St. Petersburg Branch, Special Astrophysical Observatory, 
             Russian Academy of Sciences, 196140, St. Petersburg, Russia
             }

   \date{Received September 15, 1996; accepted March 16, 1997}


  \abstract
   {Quasi-periodic pulsations (QPPs) are a common feature of solar flares, but previously there has been a lack of observational evidence to support any of the theoretical models that might explain the origin of these QPPs.}
   {We aimed to determine if there are any relationships between the QPP period and other properties of the flaring region, using the sample of flares with QPPs from \citet{2017A&A...608A.101P}. If any relationships exist then these can be compared with scaling laws for the theoretical QPP mechanisms.}
   {To obtain the flaring region properties we made use of the \emph{Atmospheric Imaging Assembly} (AIA) 1600\,\AA\ and \emph{Helioseismic and Magnetic Imager} (HMI) data. The AIA 1600\,\AA\ images allow the flare ribbons to be seen while the HMI magnetograms allow the positive and negative magnetic polarity ribbons to be distinguished and the magnetic properties determined. The ribbon properties calculated in this study were the ribbon separation distance, area, total unsigned magnetic flux, and average magnetic field strength. Only the flares that occurred within $\pm 60^{\circ}$ of the solar disk centre were included, which meant a sample of 20 flares with 22 QPP signals.}
   {Positive correlations were found between the QPP period and the ribbon properties. The strongest correlations were with the separation distance and magnetic flux. Because these ribbon properties also correlate with the flare duration, and the relationship between the QPP period and flare duration may be influenced by observational bias, we also made use of simulated data to check if artificial correlations could be introduced. These simulations show that although QPPs cannot be detected for certain combinations of QPP period and flare duration, this does not introduce an apparent correlation.}
   {There is evidence of relationships between the QPP period and flare ribbon properties, and in the future the derived scaling laws between these properties can be compared to equivalent scaling laws for theoretical QPP mechanisms.}


   \maketitle
%

\section{Introduction}

Quasi-periodic pulsations (QPPs) are frequently observed in solar and stellar flare light curves \citep[recent publications include][]{2016ApJ...830..101B, 2017JGRA..122.9841H, 2017A&A...597L...4L, 2018ApJ...858L...3K, 2018MNRAS.475.2842D, 2019MNRAS.482.5553J} and remain to be fully understood. Several different mechanisms for the generation of QPP signals have been proposed \citep[for recent reviews, see][]{2016SSRv..200...75N, 2016SoPh..291.3143V, 2018SSRv..214...45M}, but previously there has been a lack of observational evidence to support one or more of these mechanisms above the others. Limitations of the spatial and temporal resolution of solar flare data, along with saturation effects that many instruments suffer from during flares, mean that although a QPP signal may be seen well in Sun-as-a-star data, plasma motions that might be associated with the QPPs are not usually resolved in the imaging observations. Despite the lack of spatial information on the QPPs themselves, the high-quality data from instruments that have been observing the Sun over the past few years have allowed systematic statistical studies of QPPs in solar flares to be made \citep{2015SoPh..290.3625S, 2016ApJ...833..284I, 2017A&A...608A.101P, 2018SoPh..293...61D}. By examining large populations the general properties of QPPs can be determined, and these can be compared with the properties of QPPs produced by the various theoretical mechanisms.

An example of observational QPP characteristics that can be compared with theoretical mechanisms are scaling laws between the QPP and flaring region properties. For the QPP mechanisms based on standing magnetohydrodynamic (MHD) oscillations of the flaring coronal loop, the QPP period is expected to scale linearly with the loop length based on the theory of MHD modes of a straight cylinder \citep{2009SSRv..149..119N}. In some limits (such as long wavelength) this relationship might break down, however \citep{2012ApJ...761..134N}. There are other mechanisms where there may not be a linear relationship between the QPP period and loop length. For example, simulations performed by \citet{2018A&A...618A.135R}, based on the mechanism proposed by \citet{2016ApJ...833...36F}, showed oscillations with a period of 25\,s in the flare soft X-ray light curve. Based on this period, they suggested that the oscillations were most likely the result of a fast sausage mode excited by motions associated with a Kelvin-Helmholtz instability. They found that if the sausage mode were present in the whole flaring loop then the phase speed of the mode would be greater than the external Alfv\'{e}n speed, which is not consistent with MHD wave theory. Therefore they propose that the waves may be reflected before reaching the loop footpoints, due to the change of the loop cross section, and hence the associated characteristic spatial scale may be smaller than the loop length.

A linear relationship between the QPP period and an associated characteristic spatial scale is expected for some mechanisms that are not based on standing MHD oscillations \citep[e.g.][]{2016ApJ...823..150T, 2017ApJ...848..102T}, but these spatial scales may not have a linear relationship with observable spatial scales such as the loop length. \citet{2016ApJ...823..150T} found that their ``magnetic tuning fork'' mechanism produced a period that was related to the magnetic field strength as $P \propto B^{-2.1}$, and compared this to the expected scaling of $P \propto B^{-0.43}$ for QPPs produced by slow magnetoacoustic oscillations of the flaring loop. Although it is not currently possible to directly measure the magnetic field in the corona, the magnetic field strength inside a coronal loop might be expected to scale with the magnetic field measured at the loop footpoints in the photosphere.

Other mechanisms exist that would result in a QPP period independent of the flaring region properties \citep[e.g.][]{2006A&A...452..343N}, or that do not yet make testable predications of relationships between flaring region properties and the QPP period \citep[e.g.][]{2017ApJ...844....2T}.

This study continues the work of \citet{2017A&A...608A.101P}, making use of the same sample of flares with QPPs. While \citet{2017A&A...608A.101P} made the first attempt at linking QPP periods with spatial scales, the present work improves on this by focussing on the spatial scales of the flaring regions, rather than the spatial scales of the active region as a whole. In addition, we use simulated data to explore the potential for observational bias to influence correlations found between the QPP and flare properties. Details of the data used are given in Sect.~\ref{sec:obs}, while the analysis methods are described in Sect.~\ref{sec:dat}. Section~\ref{sec:res} gives the results of the search for relationships between the QPP and flare properties, along with the check for how observational bias can affect the QPP periods detected for different flare durations. Finally, conclusions are given in Sect.~\ref{sec:con}.


\section{Observations}
\label{sec:obs}

The sample used for this study is made up of flares from the long-lived active region NOAA 12172/12192/12209, which was present on the Sun between September and November 2014. During this time interval the Sun was well observed, therefore data could be taken from multiple instruments in order to give the best chance of detecting QPP signals. X-ray time series data were taken from the \emph{Geostationary Operational Environmental Satellite} (GOES) X-ray sensor (XRS), the \emph{Extreme ultraviolet SpectroPhotometer} (ESP) channel of the \emph{Extreme ultraviolet Variability Experiment} (EVE) aboard the \emph{Solar Dynamics Observatory} (SDO), the \emph{Fermi} \emph{Gamma-ray Burst Monitor} (GBM), and the \emph{Detector of the Roentgen and Gamma-ray Emissions} (DRGE) instrument aboard \emph{Vernov}. In addition, microwave time series data from \emph{Nobeyama Radioheliograph} (NoRH) were used. Characteristics of the data from these instruments, along with details of how data uncertainties were estimated for those instruments whose data did not include uncertainties, are described in \citet{2017A&A...608A.101P}.

Spatial properties of the flare ribbons were determined using data from the \emph{Atmospheric Imaging Assembly} (AIA) and \emph{Helioseismic and Magnetic Imager} (HMI), both aboard SDO. From AIA \citep{2012SoPh..275...17L} the 1600\,\AA\ images were used to identify the flare ribbons, while the HMI \citep{2012SoPh..275..207S} line-of-sight magnetogram images were used to distinguish the positive and negative magnetic polarity flare ribbons, and estimate magnetic fluxes and field strengths in the region. AIA data cubes were constructed with a cadence of 24\,s, and the time interval used covered the impulsive phase of the flare (or in other words, from the GOES flare start time to the GOES peak time). The AIA images can be subject to blooming when the CCD is saturated, which often occurs during flares, therefore frames with visible blooming were removed. For the HMI images a single frame was used for each flare, since there was little change in the magnetograms over the course of the flare impulsive phases, and these frames were taken around the times of the flare peaks. All images were used with the full resolution of 4096 $\times$ 4096 pixels.


\section{Data analysis}
\label{sec:dat}

\begin{figure*}
        \centering
        \includegraphics[width=0.9\linewidth]{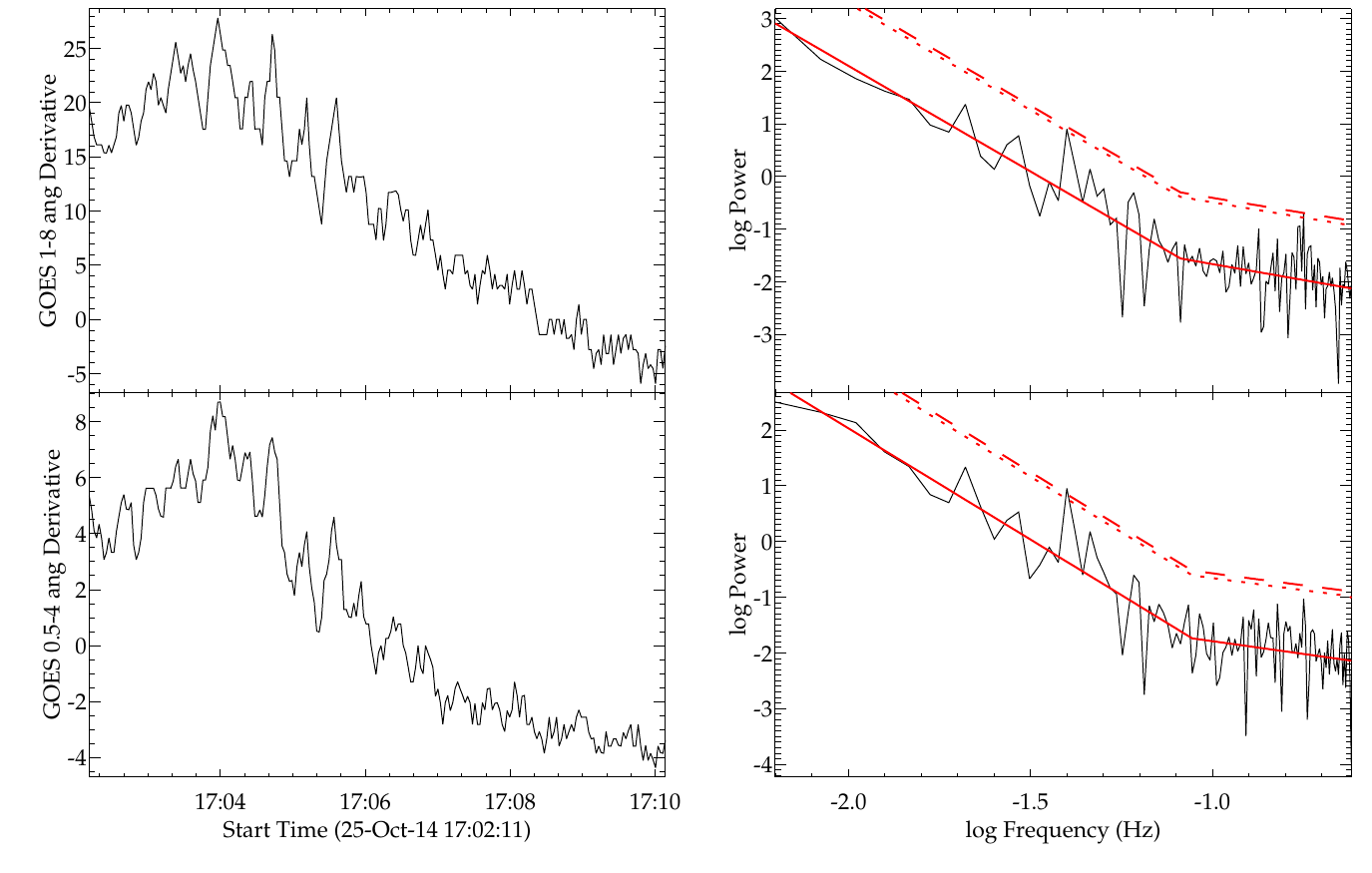}
        \caption{\emph{Left:} Time derivatives of a section of flare 092 (see Table~\ref{tab:qppflares}) observed by GOES/XRS. The top panel shows the 1--8\,\AA\ emission and the bottom panel the 0.5--4\,\AA\ emission. \emph{Right:} The corresponding power spectra, where the red solid lines are broken power-law fits to the spectra, the red dotted lines represent the 95\% confidence levels, and the red dashed lines the 99\% levels. There is a significant peak in both power spectra at a period of $25.1^{+0.7}_{-0.6}$\,s.}
        \label{fig:092goes}
\end{figure*}

\begin{table*}
\centering
\caption{List of flares with QPPs used in this study. The first column contains a numerical label for the flares (originally defined in \citealt{2017A&A...608A.101P}), the second column shows the time of the peak intensity of the flare in the GOES 1--8\,\AA\ waveband, the third column gives the GOES class, the fourth and fifth columns give the start and end times of the section of the flare where the QPP signal is most visible in the power spectrum, the sixth column is the instrument used to observe the QPP signal, and the seventh column is the QPP period. }
\label{tab:qppflares}
\begin{tabular}{c c c c c c c}	
	\hline\hline
	\rule{0pt}{10pt} Flare no. & Flare peak time (UT) & GOES class & QPP start time (UT) & QPP end time (UT) & Instrument & Period (s) \\[2.3pt]
	\hline
	\rule{0pt}{10pt} 008 & 2014-09-23 23:16:54 & M2.3 & 23:08:20 & 23:13:52 & GOES 0.5--4\,\AA	& $41.2^{+2.7}_{-2.4}$	\\[2.3pt] 
	010 & 2014-09-24 17:50:11 & C7.0 & 17:49:04 & 17:50:18 & GOES 1--8\,\AA		& $9.6^{+1.4}_{-1.1}$	\\[2.3pt]
	010 & 2014-09-24 17:50:11 & C7.0 & 17:49:01 & 17:49:50 & DRGE				& $5.8^{+0.8}_{-0.6}$	\\[2.3pt]
	049 & 2014-10-20 09:11:50 & M3.9 & 09:05:45 & 09:08:14 & GOES 0.5--4\,\AA	& $14.7^{+2.6}_{-1.9}$	\\[2.3pt]
	052 & 2014-10-20 14:43:38 & C3.1 & 14:41:47 & 14:43:33 & GOES 1--8\,\AA		& $26.1^{+3.7}_{-2.9}$	\\[2.3pt]
	054 & 2014-10-20 16:37:55 & M4.5 & 16:23:02 & 16:31:20 & GOES 1--8\,\AA		& $35.4^{+1.3}_{-1.2}$	\\[2.3pt]
	056 & 2014-10-20 19:02:46 & M1.4 & 18:57:51 & 18:59:01 & GBM 25--50\,keV	& $13.9^{+6.0}_{-3.2}$	\\[2.3pt]
	058 & 2014-10-20 22:55:58 & M1.2 & 22:45:18 & 22:49:46 & GOES 0.5--4\,\AA	& $48.4^{+10.8}_{-7.5}$	\\[2.3pt]
	068 & 2014-10-22 01:58:33 & M8.7 & 01:43:04 & 01:46:36 & GOES 0.5--4\,\AA	& $21.1^{+1.1}_{-1.0}$	\\[2.3pt]
	072 & 2014-10-22 14:28:15 & X1.6 & 14:06:56 & 14:09:30 & GOES 1--8\,\AA		& $30.3^{+3.4}_{-2.8}$	\\[2.3pt]
	072 & 2014-10-22 14:28:15 & X1.6 & 14:15:24 & 14:23:40 & GOES 0.5--4\,\AA	& $49.4^{+2.6}_{-2.4}$	\\[2.3pt]
	079 & 2014-10-24 02:43:57 & C4.2 & 02:38:30 & 02:41:20 & NoRH				& $7.9^{+0.4}_{-0.3}$	\\[2.3pt]
	081 & 2014-10-24 04:00:08 & C3.6 & 03:59:30 & 04:01:00 & NoRH				& $14.8^{+5.0}_{-3.0}$	\\[2.3pt]
	085 & 2014-10-24 21:40:30 & X3.1 & 21:19:38 & 21:23:47 & GOES 0.5--4\,\AA	& $49.6^{+5.5}_{-4.5}$	\\[2.3pt]
	092 & 2014-10-25 17:08:20 & X1.0 & 17:02:11 & 17:10:10 & GOES 0.5--4\,\AA	& $25.1^{+0.7}_{-0.6}$	\\[2.3pt]
	098 & 2014-10-26 10:56:40 & X2.0 & 10:48:52 & 10:50:34 & GBM 25--50\,keV	& $20.3^{+2.2}_{-1.9}$	\\[2.3pt]
	104 & 2014-10-26 18:15:29 & M4.2 & 18:11:18 & 18:15:24 & GOES 0.5--4\,\AA	& $20.3^{+0.9}_{-0.8}$	\\[2.3pt]
	105 & 2014-10-26 18:49:30 & M1.9 & 18:45:04 & 18:48:02 & GOES 1--8\,\AA		& $25.2^{+1.9}_{-1.7}$	\\[2.3pt]
	106 & 2014-10-26 20:21:44 & M2.4 & 20:03:42 & 20:11:18 & GOES 0.5--4\,\AA	& $36.4^{+3.2}_{-2.7}$	\\[2.3pt]
	117 & 2014-10-27 17:40:38 & M1.4 & 17:36:36 & 17:37:26 & GOES 1--8\,\AA		& $12.3^{+1.8}_{-1.4}$	\\[2.3pt]
	161 & 2014-11-16 17:48:23 & M5.7 & 17:42:46 & 17:45:24 & GOES 0.5--4\,\AA	& $19.5^{+1.3}_{-1.2}$	\\[2.3pt]
	177 & 2014-11-22 06:03:33 & C6.5 & 06:02:16 & 06:04:48 & GOES 0.5--4\,\AA	& $18.7^{+1.2}_{-1.1}$	\\[2.3pt]
	\hline
\end{tabular}
\end{table*}

The sample of flares with QPPs used in this study was taken from \citet{2017A&A...608A.101P}, who found 37 out of 181 flares had a QPP signal with a peak in the power spectrum above the 95\% global confidence level. The method used to detect the QPPs did not require any detrending of the data, and involved calculating power spectrum confidence levels that accounted for data uncertainties and the presence of red noise. Further details of the method can be found in \citep{2017A&A...602A..47P}, and an example of one of the flares with detected QPPs is shown in Fig.~\ref{fig:092goes}.

Since line-of-sight effects must be accounted for when measuring spatial properties of the flares, only flares that occurred within around $\pm 60^{\circ}$ of the solar disk could be included in this study. This gave a sample of 20 flares with 22 QPP signals, and properties of this sample are given in Table~\ref{tab:qppflares}.

\subsection{Flare ribbon properties}

\begin{figure*}
	\centering
	\includegraphics[width=0.45\linewidth]{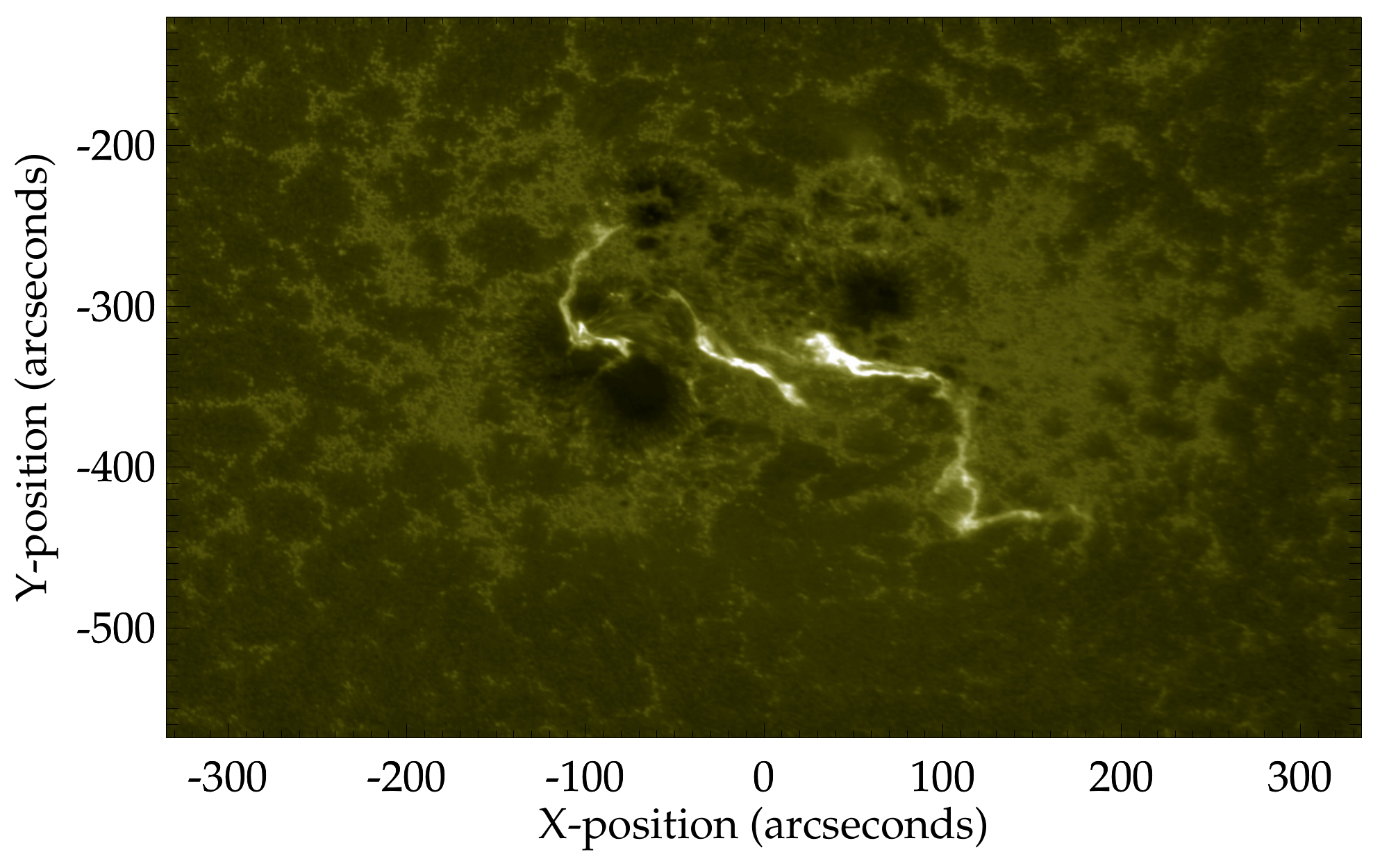}
	\hspace*{0.5cm}
	\includegraphics[width=0.45\linewidth]{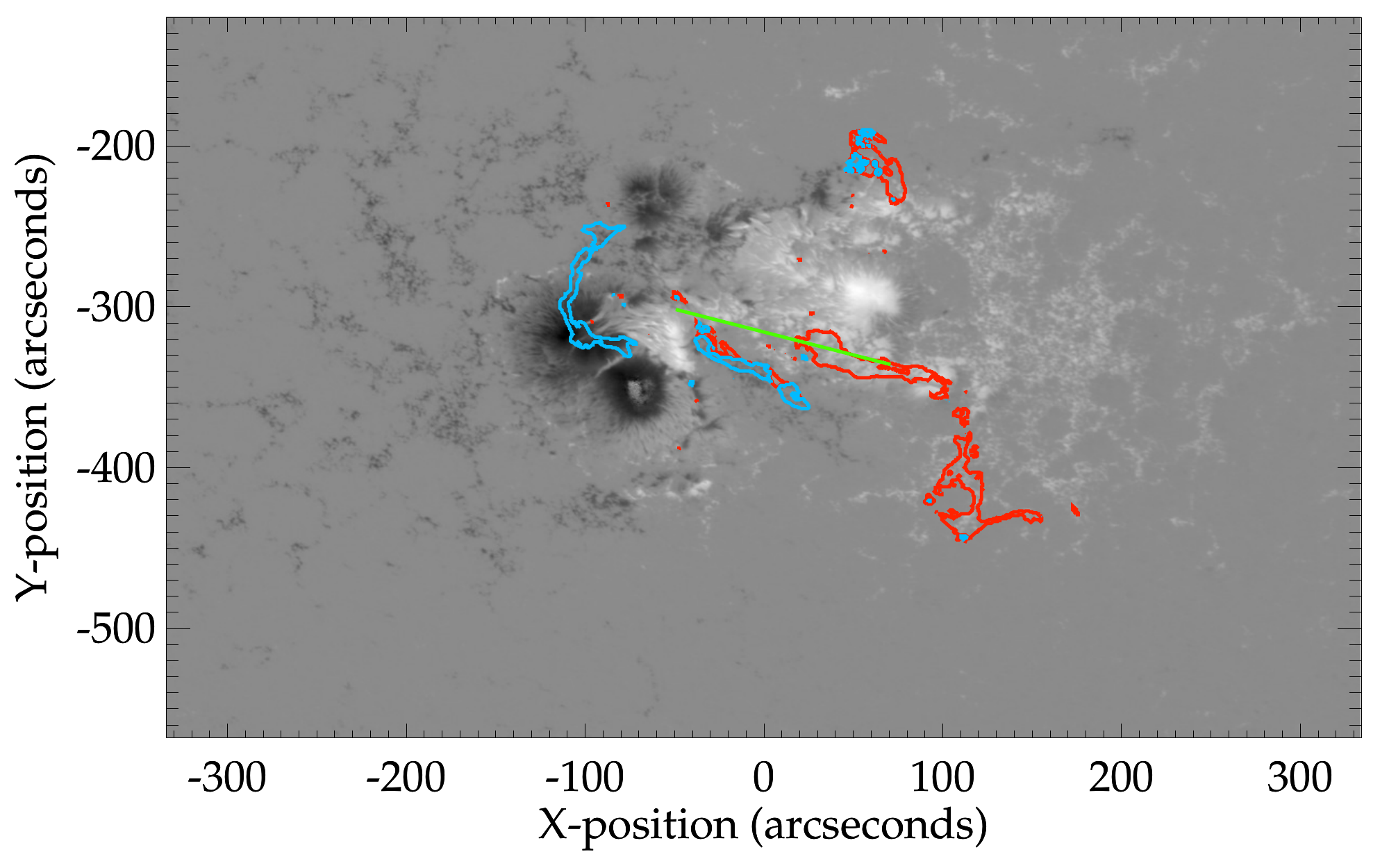}
	\caption{\emph{Left:} AIA 1600\,\AA\ images averaged over the impulsive phase of flare 092, showing full extend of the flare ribbons. \emph{Right:} HMI magnetogram showing the host active region at the time of the flare peak in the GOES 1--8\,\AA\ waveband. The red and blue contours show the positions of the composite flare ribbons with positive and negative magnetic polarity, respectively. The green line joins the geometric centroids of the positive and negative polarity ribbon components.}
	\label{fig:092ribbon}
\end{figure*}

Flare ribbon properties were determined using a similar approach to \citet{2017ApJ...834...56T}, using the AIA 1600\,\AA\ and HMI line-of-sight magnetogram data. First, for all AIA and HMI frames the line-of-sight projection effect of features closer to the solar limb appearing smaller compared to when closer to the disk centre was accounted for. This was done by differentially rotating the frames to the time when the active region was approximately at the central meridian, using the IDL Solar Software routine \emph{drot\_map}. These times were taken to be 2014 September 26 18:30:16 UT for the active region's first crossing of the solar disk as NOAA 12172, 2014 October 23 14:00:40 UT for the second crossing as NOAA 12192, and 2014 November 19 15:00:40 for the third crossing as NOAA 12209. Next, the images were cropped so that only the region containing the active region and full extent of all flare ribbons was included. The coordinates used for the cropped regions, in the form $[x_0, x_1, y_0, y_1]$ and in the units of arcseconds for the active region's three crossings of the solar disk, are $[-250, 250, -432, -147]$, $[-334, 334, -568, -120]$, and $[-551, 551, -541, 16]$, respectively.

To identify the flare ribbons, a threshold brightness for the AIA 1600\,\AA\ data was first determined. A period of time when no active regions and very few magnetic features were present on the solar disk was chosen, between 2018 March 28 21:00:00 and 21:30:00 UT. After averaging the AIA 1600\,\AA\ frames during this period and cropping the frames to the region [-550, 550, -550, 16], similar to the regions defined above, the mean and standard deviation were calculated. Based on these values, the brightness threshold was defined as the mean plus 40 times the standard deviation. For each AIA 1600\,\AA\ frame of each flare, flare ribbon pixels were defined as those with values greater than this threshold. Following the method of \citet{2017ApJ...834...56T}, a ribbon composite was then formed by selecting the pixels that exceeded the brightness threshold in any of the AIA 1600\,\AA\ frames. Doing this reduces the chance of including background noise as part of the ribbons, since a high brightness threshold is used, while ensuring that the full extent of the flare ribbons is represented. Although there is risk of overestimating the ribbon area using this approach if the ribbon separation grows with time, the flares from the active region used in this study were mostly confined and therefore do not exhibit much growth of the ribbon separation distance \citep{2015ApJ...801L..23T, 2015ApJ...804L..28S}.

To separate out the positive and negative magnetic polarity flare ribbons, for each flare the ribbon pixels were overlaid on the corresponding HMI magnetogram. The ribbons were then separated based on whether they were above a positive or negative polarity region in the HMI data. Figure~\ref{fig:092ribbon} shows an example of how the flare ribbons appear in the AIA 1600\,\AA\ data for one of the flares in Table~\ref{tab:qppflares}, along with the corresponding HMI magnetogram and the composite flare ribbon contours. HMI magnetograms and flare ribbon contours for all flares in the sample are shown in Fig.~\ref{fig:appendix}.

To calculate the flare ribbon areas, $S_{\mathrm{ribbon}}$, for each of the ribbon composite pixels the area on the solar surface that the pixel would correspond to if it were located at the disk centre was multiplied by a cosine correction factor, to account for the fact that different pixels correspond to different surface areas due to the spherical nature of the Sun. These corrected pixel areas were then summed together to obtain the total area of the flare ribbons. The ribbon separation distance, $d_{\mathrm{ribbon}}$, was estimated from the great circle distance between the geometric centroids of the positive and negative polarity components of the flare ribbons. The total unsigned magnetic flux, $|\Phi|_{\mathrm{ribbon}}$, below the ribbons was defined as $|\Phi|_{\mathrm{ribbon}} = \int_{S_{\mathrm{ribbon}}} |B| \,\mathrm{d}s$, and the average magnetic field strength as $|B|_{\mathrm{ribbon}} = |\Phi|_{\mathrm{ribbon}}/S_{\mathrm{ribbon}}$. The calculated flare ribbon properties are given in Table~\ref{tab:appendix}.

\subsection{Simulated flare light curves}
\label{sec:sim}

In order to explore the potential for observational bias to result in apparent relationships between the QPP periods and flare properties in the observations, a set of simulated flare light curves was generated. For these simulated flares 16 flare durations, QPP periods, and QPP durations were used, which were all uniformly distributed in log space. This meant a total of 4096 simulated flares were created, each with a different combination of the 16 QPP periods, QPP durations, and flare durations. The flares were given a time cadence of 1\,s, with log durations evenly spaced between 1.300 and 3.505 (or between around 20 and 3200\,s in linear space), log QPP periods between 0.7 and 1.9 (or 5 and 79\,s in linear space), and log QPP durations between 1.0 and 2.5 (or 10 and 316\,s in linear space). These parameter spaces were chosen to be similar to the observed flare and QPP parameters reported by \citet{2017A&A...608A.101P}. For the basic flare time profile we made use of the following expression from \citet{2017SoPh..292...77G}:
\begin{equation}
	I(t) = \frac{1}{2}\sqrt{\pi}AC\exp \Bigg[ D\,(B-t) + \frac{C^2D^2}{4}\Bigg] \Bigg[ \mathrm{erf} (H) - \mathrm{erf} \left( H - \frac{t}{C}\right) \Bigg] \,,
\label{eq:flareprofile}
\end{equation}
where
\begin{equation}
	H = \frac{2B + C^2D}{2C} \,,
\end{equation}
$I(t)$ is the flare intensity as a function of time, and $A$, $B$, $C$, $D$ are arbitrary parameters. For this set of simulated flares, these arbitrary parameters were set to: $A=1$, $B = t_{\mathrm{flare}}/15$, $C = t_{\mathrm{flare}}/10$, and $D = 3/t_{\mathrm{flare}}$.

The QPP signal was modelled as a Gaussian modulated sinusoid, as described by the following expression:
\begin{equation}
	I_{\mathrm{QPP}}(t) = A_{\mathrm{QPP}}\cos\left(\frac{2\pi t}{P}\right) \exp\left(\frac{-(t-t_0)^2}{\tau_{\mathrm{G}}^2/2}\right) \,,
\label{eq:qppprofile}
\end{equation}
where $I_{\mathrm{QPP}}(t)$ is the QPP signal intensity as a function of time, $A_{\mathrm{QPP}}$ is the signal amplitude (set to 0.1 for all flares), $P$ is the period, $t_0$ is set to be the flare peak time, and $\tau_{\mathrm{G}}$ is the Gaussian decay time.

\begin{figure}
	\centering
	\includegraphics[width=0.9\linewidth]{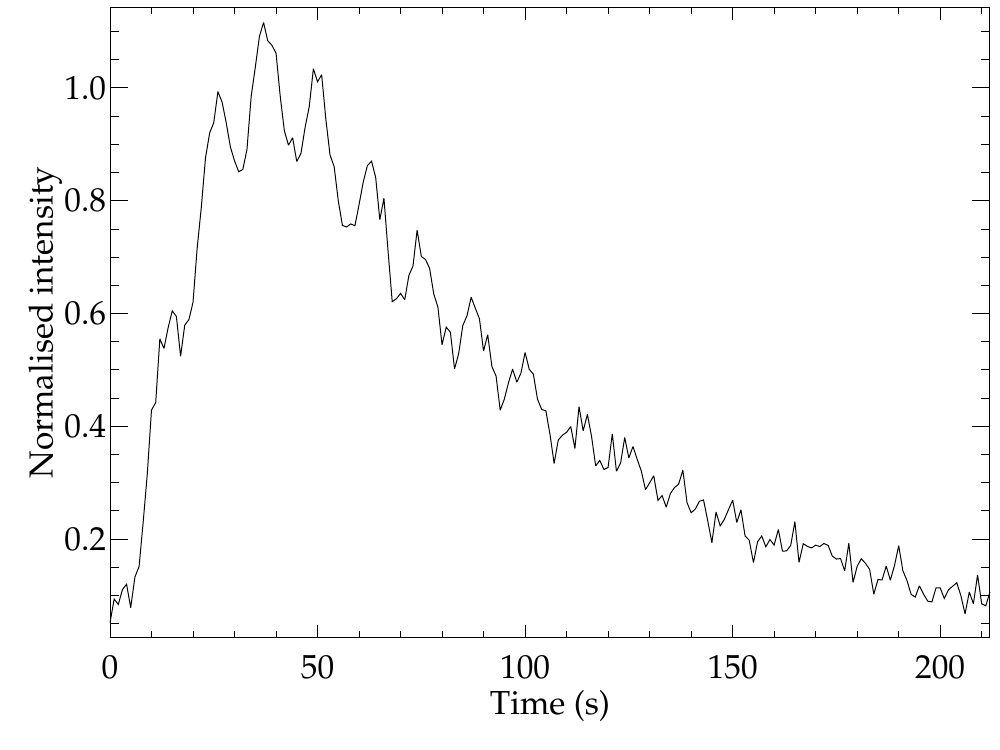}
	\caption{One of the simulated flare light curves, with a flare duration of 213\,s, a QPP period of 12\,s and a QPP decay time of 100\,s.}
	\label{fig:example}
\end{figure}

After normalising Eq.~(\ref{eq:flareprofile}) by the peak intensity of the flare, so that the resulting flare peak intensity was 1, it was summed with Eq.~(\ref{eq:qppprofile}) and normally distributed white noise with a standard deviation of 0.02. One of the simulated flares is shown in Fig.~\ref{fig:example} as an example.

Once the set of simulated flares had been generated, we attempted to detect the QPPs using a simplified version of the method used in \citet{2017A&A...608A.101P}, since there were no uncertainties on the simulated data. For each flare a power spectrum was first generated, and a broken power-law model with the following form was fitted to the spectrum:
\begin{equation}
	\log\left[\hat{\mathcal{P}}(f)\right] = 
		\begin{cases}
			-\alpha\log\left[f\right] + c & \text{if } f < f_{\mathrm{break}} \\
			-\left(\alpha - \beta\right)\log\left[f_{\mathrm{break}}\right] - \beta\log\left[f\right] + c & \text{if } f > f_{\mathrm{break}}\,,
		\end{cases}
\end{equation}
where $\hat{\mathcal{P}}(f)$ is the spectral power of the fit as a function of frequency, $f_{\mathrm{break}}$ is the frequency at which the power law break occurs, $\alpha$ and $\beta$ are power law indices and $c$ is a constant.

Next the confidence levels were calculated. Noise in the power spectrum follows a chi-squared, two degrees of freedom (d.o.f.) distribution, and the corresponding probability density is
\begin{equation}
	p(z) = \mathrm{e}^{-z}\,.
\end{equation}
Therefore the probability of having a value $Z$ in the power spectrum that is greater than some threshold $z$ is
\begin{equation}
	\text{Pr}\left\{Z>z\right\} = \int_{z}^{\infty} \mathrm{e}^{-z'} \mathrm{d}z' = e^{-z}\,.
\label{eq:prob1}
\end{equation}
In order to calculate the `global' confidence level the number of data points in the spectrum must be accounted for. For a power spectrum sampled at $N$ independent frequencies the probability is equivalent to:
\begin{equation}
	\text{Pr}\left\{Z>z\right\} = 1 - \left(1 - \epsilon _{N}\right)^{1/N} \approx \epsilon _{N}/N\,,
\label{eq:prob2}
\end{equation}
where $\epsilon _{N}$ is the false alarm probability, and the approximation holds when is $\epsilon _{N}$ small \citep{2002MNRAS.336..979C}. Equating Eq.~(\ref{eq:prob1}) and \ref{eq:prob2} and rearranging gives
\begin{equation}
	z \approx \ln\left(\frac{N}{\epsilon _{N}}\right)\,.
\end{equation}
Finally, the fact that the power spectrum is not normalised must be accounted for, since the above expression will only give the confidence level for a power spectrum dominated by white noise and with a mean value of 1. For the flare power spectra the noise is distributed around the broken power law, so the confidence level can be found from $\log[\hat{\mathcal{P}}_j] + \log[z\langle J_j / \hat{\mathcal{P}}_j\rangle]$, where $J_j$ is the observed spectral power at frequency $f_j$. The 95\% global confidence level (corresponding to $\epsilon _{N} = 0.05$) was used as the detection threshold for the QPP signals. In addition, only significant spectral peaks with periods greater than four times the time cadence and less than a quarter of the duration of the time series were included, since outside of this range it is unclear that a periodic signal can be reliably detected, and the same constraint was applied in \citet{2017A&A...608A.101P}.


\section{Results and discussion}
\label{sec:res}

\subsection{Comparing QPP and flare properties}

\begin{figure}
	\centering
	\includegraphics[width=0.9\linewidth]{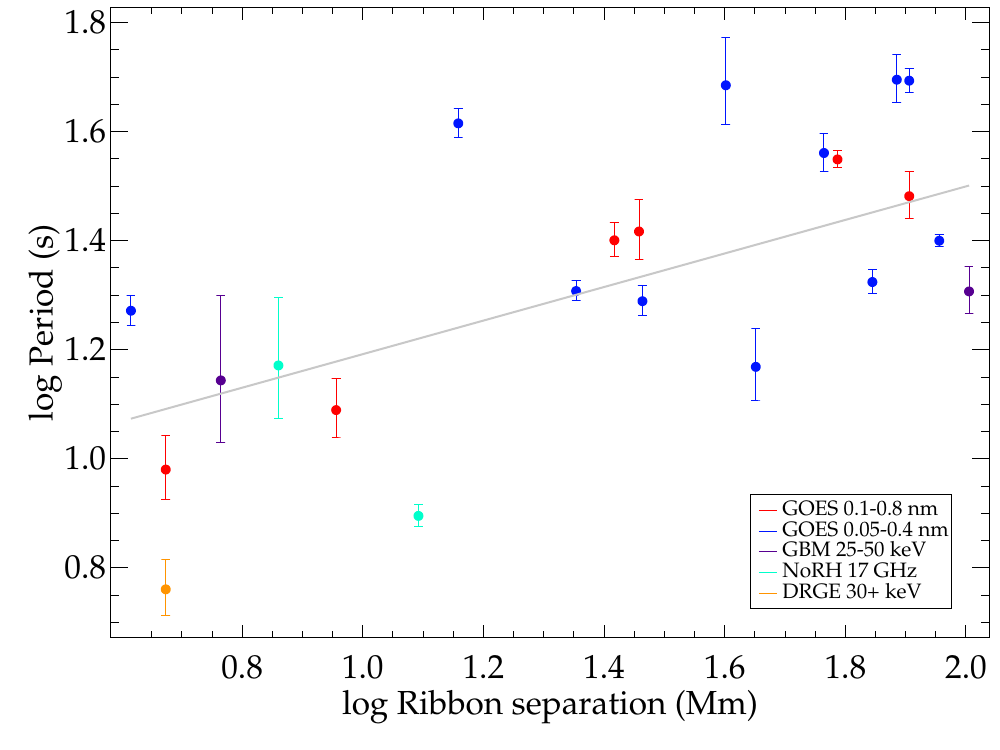}
	\caption{QPP periods plotted against the separation of the geometric centroids of the two flare ribbons. The different coloured points correspond to the different instruments used to detect the QPP signals, and the grey line is a linear fit in log space. The Pearson correlation is 0.53 with a p-value of 0.01, and the Spearman correlation is 0.64 with a p-value of 0.001.}
	\label{fig:sep}
\end{figure}

\begin{figure}
	\centering
	\includegraphics[width=0.9\linewidth]{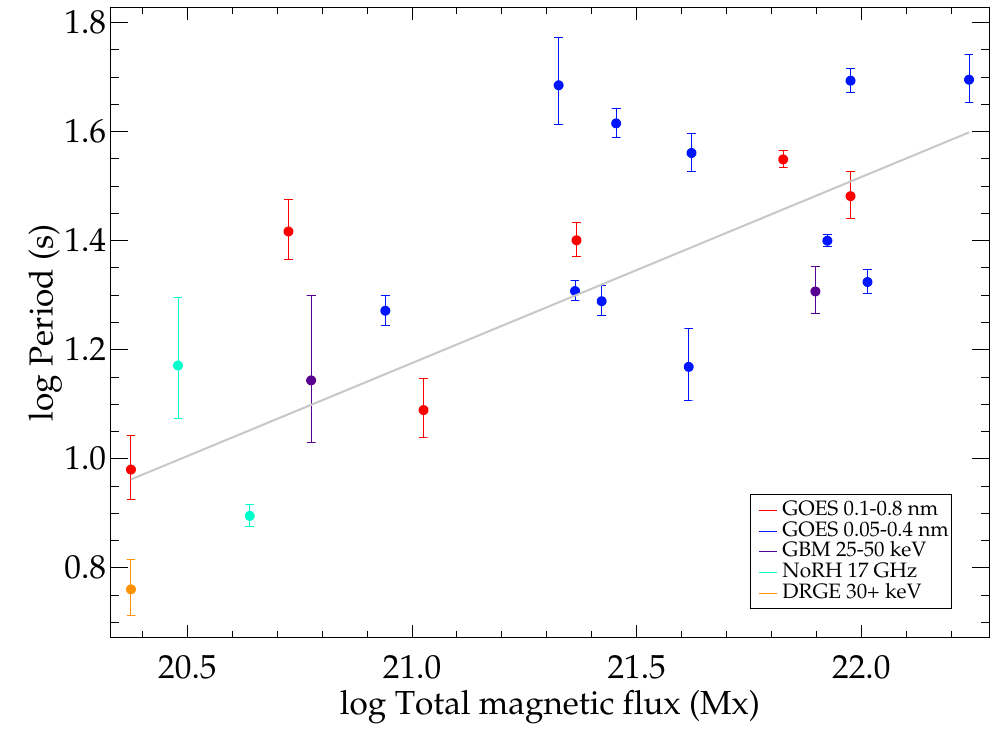}
	\caption{QPP periods plotted against the total unsigned magnetic flux of the photosphere below the flare ribbons. The Pearson correlation is 0.59 with a p-value of 0.004, and the Spearman correlation is 0.68 with a p-value of 0.0005.}
	\label{fig:mflux}
\end{figure}

\begin{figure}
	\centering
	\includegraphics[width=0.9\linewidth]{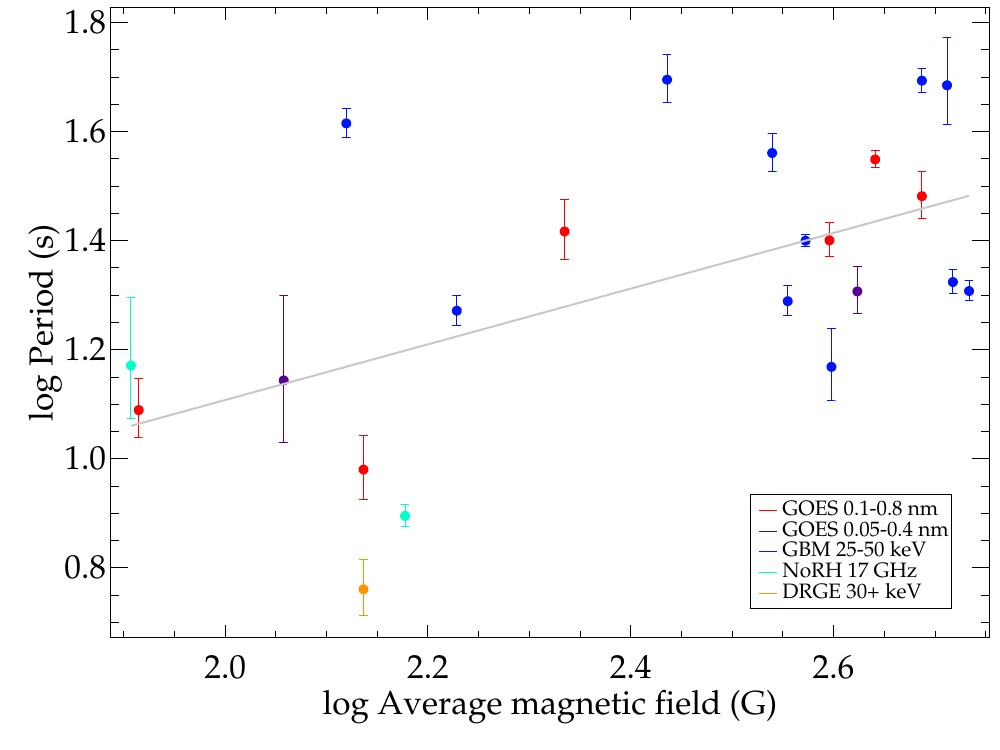}
	\caption{QPP periods plotted against the average magnetic field strength of the photosphere below the flare ribbons. The Pearson correlation is 0.47 with a p-value of 0.03, and the Spearman correlation is 0.50 with a p-value of 0.02.}
	\label{fig:mag}
\end{figure}

\begin{figure}
	\centering
	\includegraphics[width=0.9\linewidth]{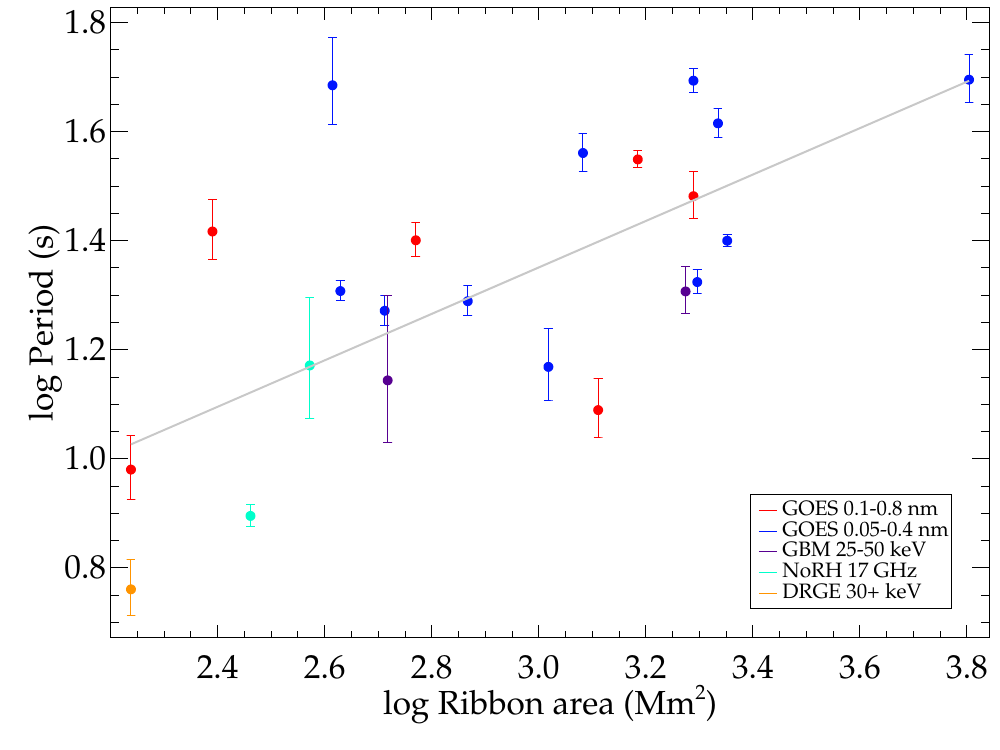}
	\caption{QPP periods plotted against the area of the flare ribbon. The Pearson correlation is 0.58 with a p-value of 0.005, and the Spearman correlation is 0.59 with a p-value of 0.004.}
	\label{fig:area}
\end{figure}

Figures \ref{fig:sep}--\ref{fig:area} show scatter plots of the detected QPP period, $P$, with the flare ribbon separation distance $d_{\mathrm{ribbon}}$, total unsigned magnetic flux $|\Phi|_{\mathrm{ribbon}}$, average magnetic field strength $|B|_{\mathrm{ribbon}}$, and area $S_{\mathrm{ribbon}}$, respectively. All plots show evidence of a positive correlation, and power-law fits give the following relationships:
\begin{equation}
	\log P = (0.31 \pm 0.01)\log d_{\mathrm{ribbon}} + (0.88 \pm 0.02)\,,
	\label{eq:p_vs_d}
\end{equation}
\begin{equation}
	\log P = (0.34 \pm 0.01)\log |\Phi|_{\mathrm{ribbon}} + (-6.0 \pm 0.3)\,,
	\label{eq:p_vs_phi}
\end{equation}
\begin{equation}
	\log P = (0.51 \pm 0.03)\log |B|_{\mathrm{ribbon}} + (0.09 \pm 0.07)\,,
	\label{eq:p_vs_B}
\end{equation}
\begin{equation}
	\log P = (0.42 \pm 0.02)\log S_{\mathrm{ribbon}} + (0.07 \pm 0.05)\,.
\end{equation}
Since these relationships are all non-linear, the Spearman's Rank correlation coefficient is a more suitable measure of the correlation than the Pearson correlation coefficient if the properties are considered in linear space. Based on this, the strongest correlations are with the total unsigned magnetic flux (with a Spearman correlation of 0.68) and the ribbon separation distance (with a Spearman correlation of 0.64). If the Pearson correlation coefficients are calculated based on the logarithm of the properties, then values of 0.68, 0.73, 0.57, and 0.65 are obtained for Figs.~\ref{fig:sep}--\ref{fig:area}, respectively.

\begin{figure*}
	\centering
	\includegraphics[width=0.45\linewidth]{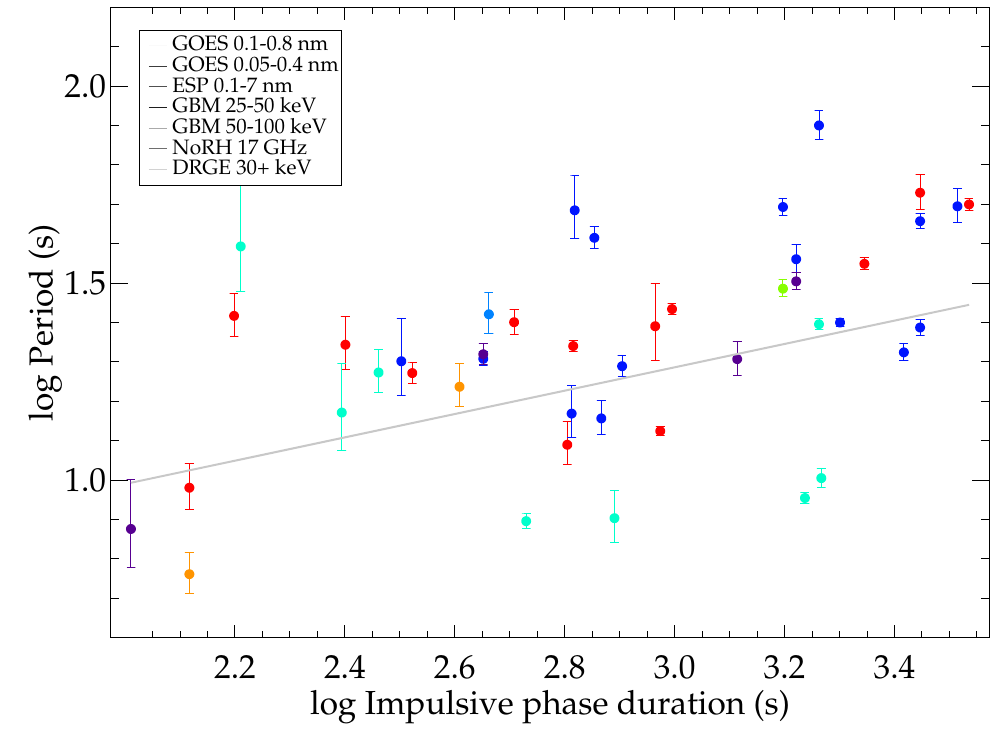}
	\hspace*{0.5cm}
	\includegraphics[width=0.45\linewidth]{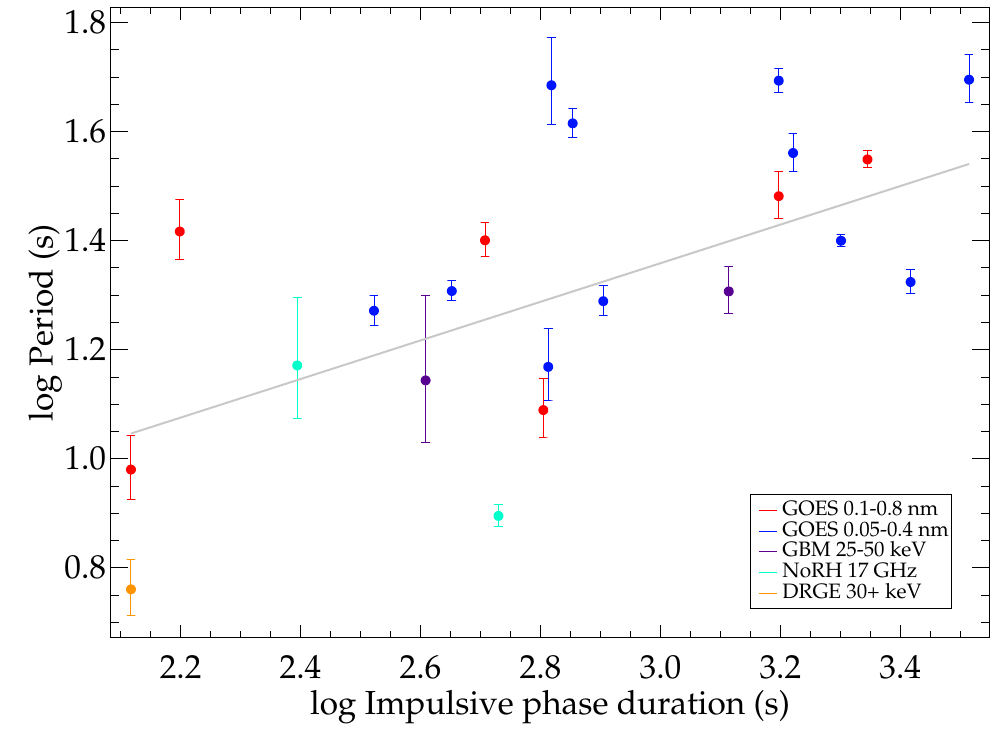}
	\caption{Two variants of the QPP period plotted against the flare impulsive phase duration. \emph{Left:} This plot includes the full set of 37 flares with QPPs found by \citet{2017A&A...608A.101P}. The Pearson correlation is 0.53 with a p-value of 0.0002, and the Spearman correlation is 0.50 with a p-value of 0.0006. \emph{Right:} This plot includes only those 20 flares which occurred within $\pm 60^{\circ}$ of the solar disk centre, and hence is the same sample of flares used in Figs.~\ref{fig:sep}--\ref{fig:area}. The Pearson correlation coefficient is 0.56 with a p-value of 0.006, and the Spearman correlation coefficient is 0.65 with a p-value of 0.001.}
	\label{fig:dur}
\end{figure*}

\begin{figure*}
	\centering
	\includegraphics[width=0.3\linewidth]{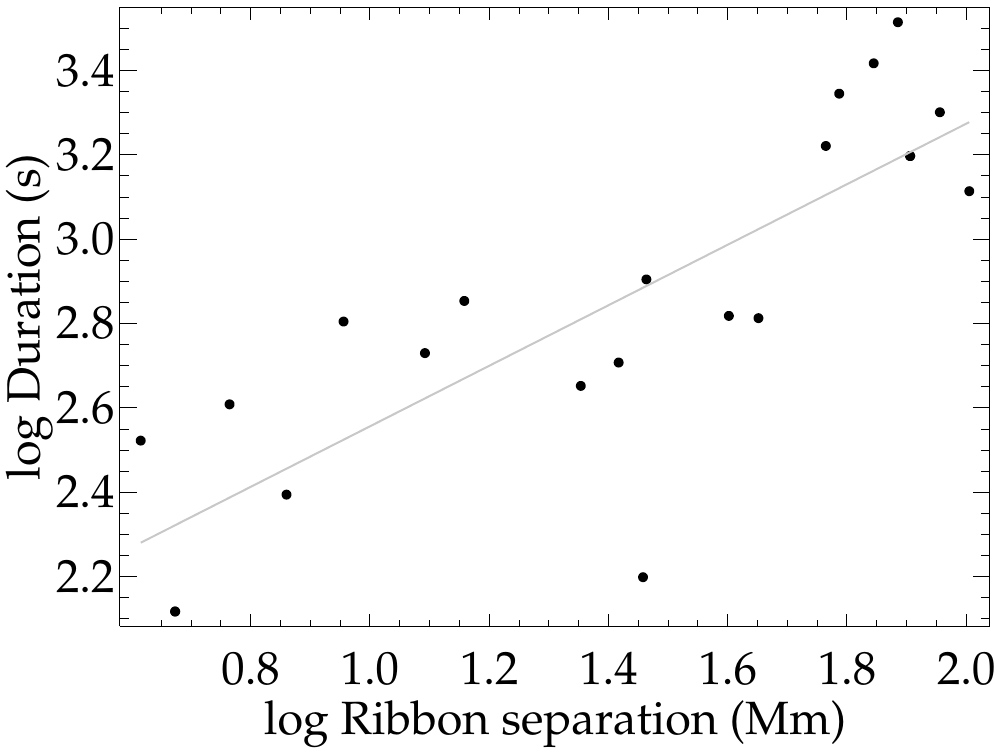}
	\hspace*{0.3cm}
	\includegraphics[width=0.3\linewidth]{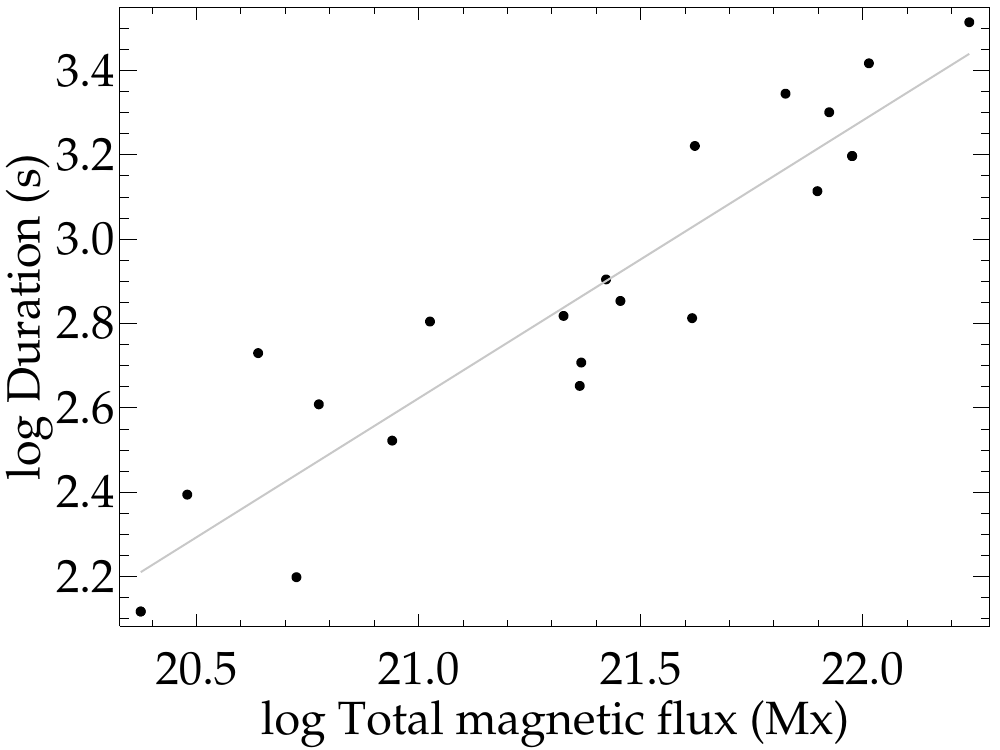}
	\hspace*{0.3cm}
	\includegraphics[width=0.3\linewidth]{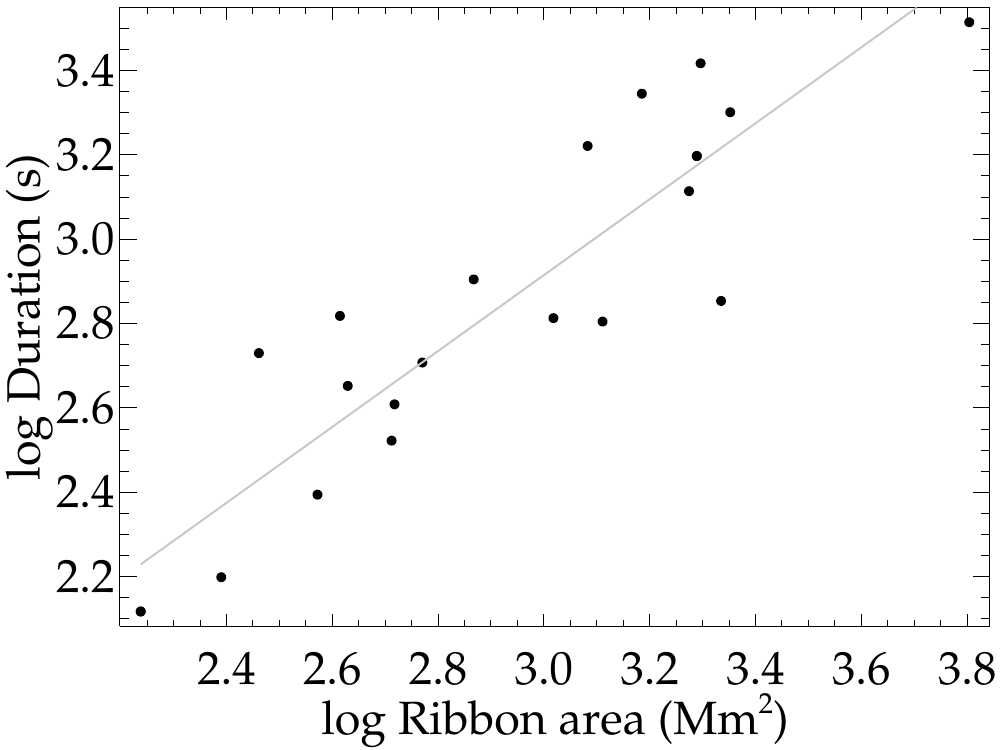}
	\caption{Flare impulsive phase durations plotted against the flare ribbon separation distance (\emph{left}), total magnetic flux (\emph{middle}), and area (\emph{right}). The grey lines show linear fits in log space. \emph{Left:} The Pearson correlation is 0.80 with a p-value of $9 \times 10^{-6}$, and the Spearman correlation is 0.84 with a p-value of $1 \times 10^{-6}$. \emph{Middle:} The Pearson correlation is 0.93 with a p-value of $5 \times 10^{-10}$, and the Spearman correlation is 0.92 with a p-value of $1 \times 10^{-9}$. \emph{Right:} The Pearson correlation is 0.83 with a p-value of $1 \times 10^{-6}$, and the Spearman correlation is 0.88 with a p-value of $9 \times 10^{-8}$.}
	\label{fig:ribbon_vs_dur}
\end{figure*}

\citet{2017A&A...608A.101P} also found a correlation between the QPP period and the total duration of the flares as measured in the GOES 1--8\,\AA\ waveband. Since it is sometimes difficult to determine exactly when a flare ends in the GOES/XRS data, a more reliable estimate of the flare duration may be the time between when the GOES 1--8\,\AA\ flux starts to increase and when the 1--8\,\AA\ flux reaches its peak, which roughly corresponds to the impulsive phase of the flare. Therefore plots of the QPP period against the impulsive phase duration of the flare, $t_{\mathrm{flare}}$, are shown in Fig.~\ref{fig:dur}, where the left plot is for the full sample of flares with QPPs identified by \citet{2017A&A...608A.101P} and the right plot is for the subset of flares used in this study. These still show positive correlations, and power-law fits to the two versions of the QPP period versus flare duration plots gave the following relationships:
\begin{equation}
	\log P = (0.30 \pm 0.01)\log t_{\mathrm{flare}} + (0.40 \pm 0.02)\,,
\end{equation}
\begin{equation}
	\log P = (0.35 \pm 0.02)\log t_{\mathrm{flare}} + (0.30 \pm 0.05)\,.
\end{equation}
There is potential for observational bias to limit which QPP signals can be detected, as mentioned by \citet{2017A&A...608A.101P}, since for long-period QPPs in a short-duration flare there would not be a sufficient number of flare intensity pulses present for the signal to have a significant peak in the power spectrum. \citet{2017ApJ...834...56T} showed that the flare ribbon properties correlate strongly with the flare duration, hence if there is potential for observational bias to affect the relationship between the QPP period and flare duration, then there is also potential for the relationships between the QPP period and ribbon properties to be affected. Figure~\ref{fig:ribbon_vs_dur} shows plots similar to those presented by \citet{2017ApJ...834...56T} but for the flare sample used in this study, and power-law fits to the plots give the following relationships: 
\begin{equation}
	\log t_{\mathrm{flare}} = (0.7 \pm 0.1)\log d_{\mathrm{ribbon}} + (1.8 \pm 0.2)\,,
\end{equation}
\begin{equation}
	\log t_{\mathrm{flare}} = (0.65 \pm 0.06)\log |\Phi|_{\mathrm{ribbon}} + (-11 \pm 1)\,,
\end{equation}
\begin{equation}
	\log t_{\mathrm{flare}} = (0.90 \pm 0.09)\log S_{\mathrm{ribbon}} + (0.2 \pm 0.3)\,.
\end{equation}
Despite using a different estimate for the flare duration, these relationships and the correlation coefficients given in Fig.~\ref{fig:ribbon_vs_dur} are consistent with those reported by \citet{2017ApJ...834...56T}. The results of \citet{2017ApJ...834...56T} are also consistent with other studies, that show that the flare duration is correlated with the GOES peak X-ray flux \citep{2002A&A...382.1070V}, and the peak X-ray flux is correlated with the flare ribbon area \citep{2017ApJ...845...49K}.

\subsection{Checking for observational biases using simulated data}

\begin{figure*}
	\centering
	\includegraphics[width=0.45\linewidth]{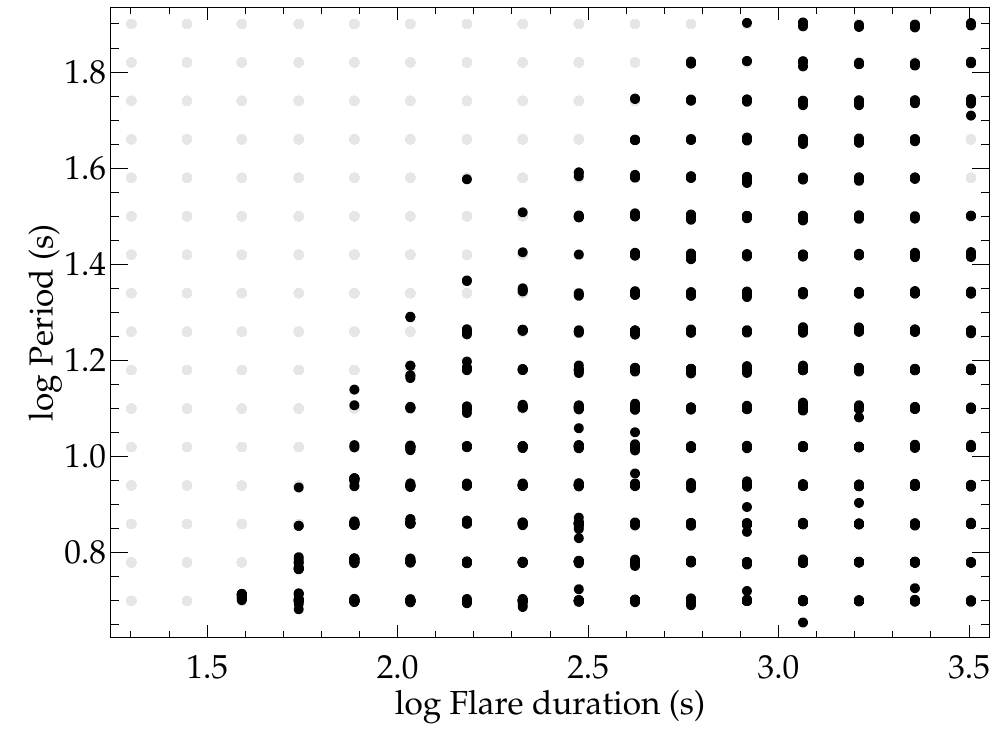}
	\hspace*{0.5cm}
	\includegraphics[width=0.45\linewidth]{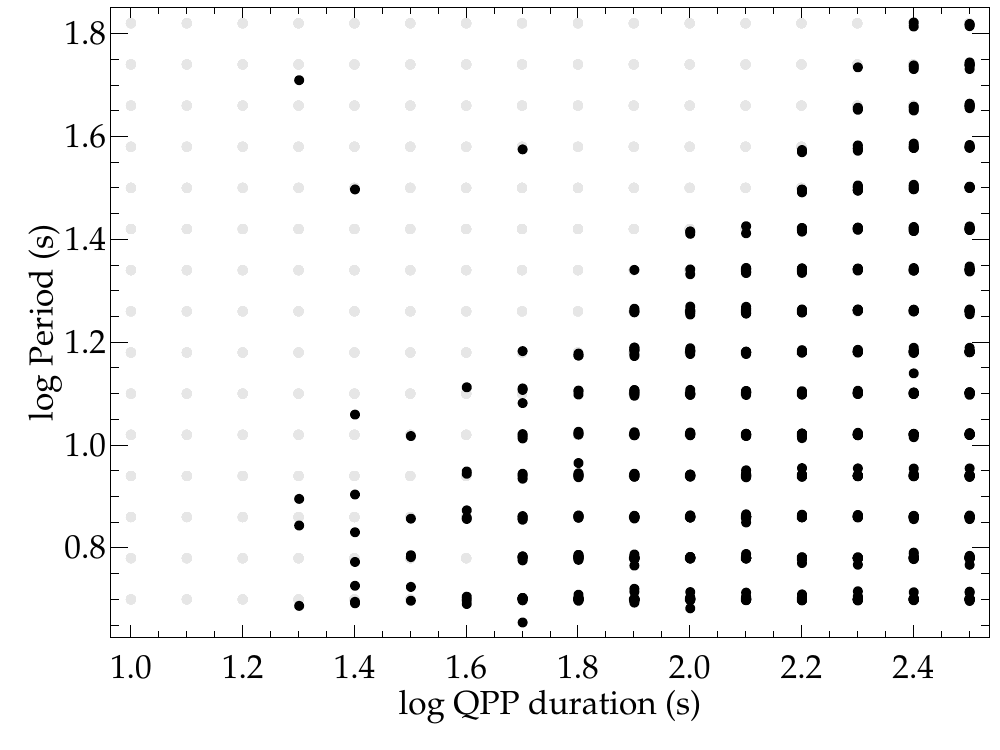}
	\caption{Simulated flare QPP period plotted against the flare duration (\emph{left}) and QPP signal duration (\emph{right}). The grey points represent the input parameters of the simulated flares with QPPs, while the black points represent the simulated flares with QPP signals that were detected using the method described in Sect.~\ref{sec:sim}. \emph{Left:} The Pearson correlation is 0.23 with a p-value of 0.0, and the Spearman correlation is 0.30 with a p-value of $4 \times 10^{-27}$. \emph{Right:} The Pearson correlation is 0.31 with a p-value of 0.0, and the Spearman correlation is 0.30 with a p-value of $3 \times 10^{-27}$.}
	\label{fig:sim1}
\end{figure*}

Figure~\ref{fig:sim1} shows results from the first set of simulated data, described in Sect.~\ref{sec:sim}. The parameters input into the model flares are represented by the grey points, while the black points represent the parameters of the flares with QPP signals that were detected by the method described in Sect.~\ref{sec:sim}. The majority of detections made match the uniform in log-space sampling of periods and durations that were input into the simulated flares (in other words, the black points mostly lie above the grey points), but a few inaccurate detections can be seen, where the black points are between the grey points. This is due to using the 95\% confidence level as the detection threshold. The important aspect of these plots is the distribution of the black points. As expected, observational bias has resulted in triangular shapes to the plots in Fig.~\ref{fig:sim1}, with an absence of points in the long-period, short-duration regions. There is no absence of points in the short-period, long-duration regions, however, hence the correlation coefficients are small.

\begin{figure}
	\centering
	\includegraphics[width=0.9\linewidth]{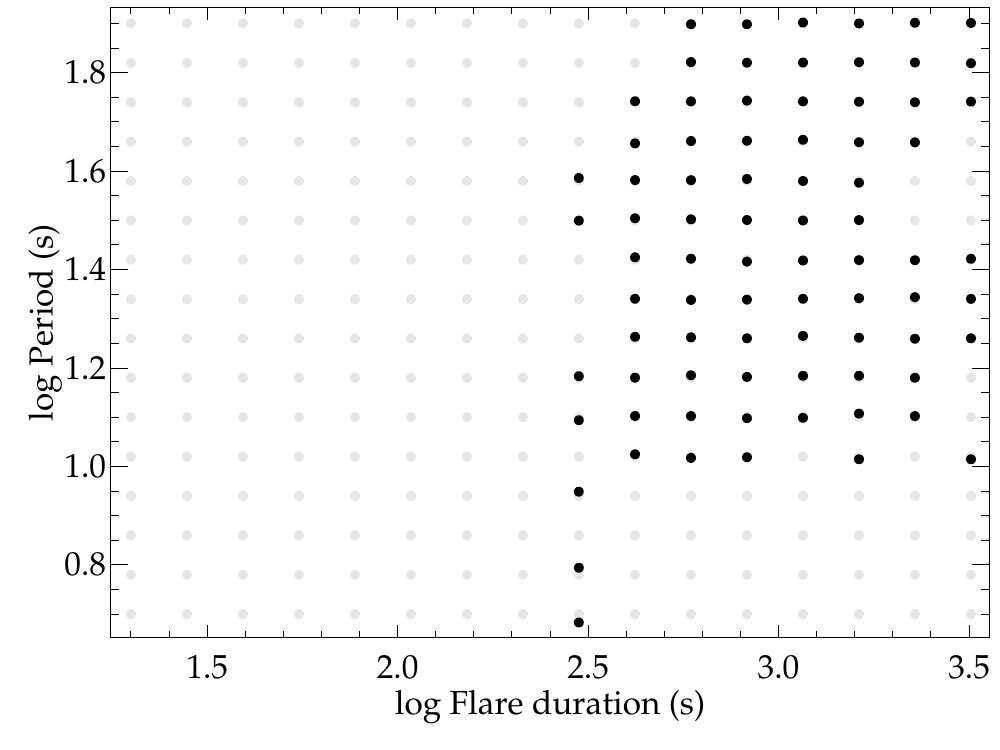}
	\caption{Simulated flare QPP period plotted against the flare duration for the case where the QPP signal duration is set to depend on the flare duration. The grey points represent the input parameters of the simulated flares with QPPs, while the black points represent the simulated flares with QPP signals that were detected using the method described in Sect.~\ref{sec:sim}. The Pearson correlation is 0.18 with a p-value of 0.1, and the Spearman correlation is 0.21 with a p-value of 0.07.}
	\label{fig:sim2}
\end{figure}

\citet{2017A&A...608A.101P} showed a positive correlation between the QPP period and the duration of the QPP signal, defining the duration as the time interval that gave the most significant peak in the power spectrum. Fitting a linear model gave the relationship:
\begin{equation}
	\log P = (0.62 \pm 0.03)\log \tau - (0.07 \pm 0.07),
\label{eq:p_vs_tau}
\end{equation}
where $P$ is the period and $\tau$ is the QPP signal duration time. To check if this relationship could result in an apparent dependence between the period and flare duration, another set of simulated flares was generated assuming the relationship to be real. For this set of flares the same ranges of periods and flare durations were used as before, but the QPP duration was instead set to depend on the period (rearranging Eq.~(\ref{eq:p_vs_tau}) to give $\tau = 1.30P^{1.61}$). Since a Gaussian decay time was used for the QPPs in the simulated flares, whereas for the set of real flares the QPP duration was taken to be the time interval that gave the most prominent peak in the power spectrum, here we assume $\tau \approx 2\tau_{\mathrm{G}}$. Attempting to detect the QPPs in the simulated flares, and then producing a period versus flare duration plot for the flares with detected QPPs, gives the plot shown in Fig.~\ref{fig:sim2}. The correlation coefficients for this plot are fairly small, therefore this is evidence against the possibility that observational bias and a relationship between the QPP period and QPP duration could result in an apparent relationship between the QPP period and flare duration, with no physical basis behind the relationship. Hence the results of analysing these simulated data support the idea that the observed QPP period versus flare duration relationship is, at least in part, a true relationship. 

\section{Conclusions}
\label{sec:con}

In this study a comparative analysis of QPPs and properties of flare ribbons in the long-lived super-active region NOAA 12172/12192/12209 has been performed. An important feature of this active region is that the hosted flares lack a substantial increase of the ribbon separation distance, thus making the process of determining ribbon properties more straight-forward.

This study has revealed correlations between the QPP period and all four flare ribbon properties, determined using the AIA 1600\,\AA\ and HMI data, as well as the duration of the flare impulsive phase. These ribbon properties are the area, separation distance, total unsigned magnetic flux, and average magnetic field strength. Since there is potential for observational bias to affect the relationship observed between the QPP period and flare duration, and the flare duration strongly correlates with the flare ribbon properties, tests with simulated flare light curves were performed. These suggest that the correlations cannot be explained by observational bias alone, and therefore support the idea that they are real. In addition, the strongest correlation is between the QPP period and total unsigned magnetic flux in the flare ribbons, not the flare duration. 

The results of this study open up the possibility for revealing the mechanisms behind QPPs. The obtained empirical scaling laws of the QPP period with the flare parameters do not seem to match any of the current proposed theoretical mechanisms that have comparable scaling laws, however. For example, on one hand the positive correlation of the period with the ribbon separation distance is consistent with a similar property of standing kink oscillations \citep{2016A&A...585A.137G}. On the other hand, the observed increase of the period with the magnetic field strength is not typical of kink oscillations; the increase of the field strength should lead to an increase of the Alfv\'{e}n and kink speeds, and hence a decrease of the period. Likewise, the period of sausage oscillations decreases as the fast magnetoacoustic speed (and hence the magnetic field) increases, while the period is only weakly dependent on the oscillating loop length (which would relate to the ribbon separation) \citep[e.g.][]{2012ApJ...761..134N}. Kink and sausage oscillations should not be disregarded as possible mechanisms for QPPs, however, as both kink and sausage oscillation periods are also determined by the mass of the oscillating loop, which may affect the scaling. In addition, for some promising mechanisms the relationships between the QPP period and the flaring region parameters have not yet been addressed. This is especially applicable to the mechanisms based on repetitive spontaneous magnetic reconnection (see \citet{2016SoPh..291.3143V, 2018SSRv..214...45M} for recent reviews). Thus, the empirically determined scaling presented in this paper demonstrates the need for dedicated theoretical modelling of this scaling for various QPP mechanisms, to allow their validation.

In addition, while this study is the first successful attempt at observing these scaling laws, the obtained scaling laws are not yet definitive and some possible shortcomings should be considered. Firstly a larger sample size should be used to improve the accuracy and precision of the scaling laws. Secondly, since the QPP detection method used is based on the periodogram, only stationary and weakly non-stationary oscillatory signals in the time series data could be detected. The effects of strong modulation of the QPP amplitude or period were not accounted for, therefore the detection method is conservative (see \citealt{2017A&A...608A.101P} for a discussion). Another possible shortcoming is the lack of any classification of the QPPs, the need for which has been stressed in \citet{2019PPCF...61a4024N}. QPPs of different classes may be caused by different mechanisms, that correspond to different scaling laws between the period and flare parameters. The identification of different classes of QPPs would allow us to search for scaling laws typical for the different QPP classes, and verify the hypotheses of their mechanisms. Nevertheless, we consider the obtained scaling laws as encouraging findings that should stimulate development of the theoretical models and further observational studies.


\begin{acknowledgements}
The authors would like to acknowledge the ISSI International Team "Quasi-periodic Pulsations in Stellar Flares: a Tool for Studying the Solar-Stellar Connection" led by A-MB for many productive discussions, and are grateful to I. N. Myagkova and A. V. Bogomolov for providing the data from \emph{Vernov}, along with the GOES/XRS, SDO/EVE, SDO/HMI, SDO/AIA, \emph{Fermi}/GBM, and Nobeyama teams for providing the rest of the data used. A-MB acknowledges support from the Royal Society International Exchanges grant IEC\textbackslash R2\textbackslash 170056, and VMN acknowledges support from the STFC consolidated grant ST/P000320/1 and the Russian Foundation for Basic Research grant No. 17-52-80064 BRICS-A. Flare information was provided courtesy of SolarMonitor.org.
\end{acknowledgements}


\bibliographystyle{aa}
\bibliography{ms}

\onecolumn

\begin{appendix}

\section{Additional figures}
\label{sec:figs}

\begin{figure}[!h]
	\centering
	\includegraphics[width=0.38\linewidth]{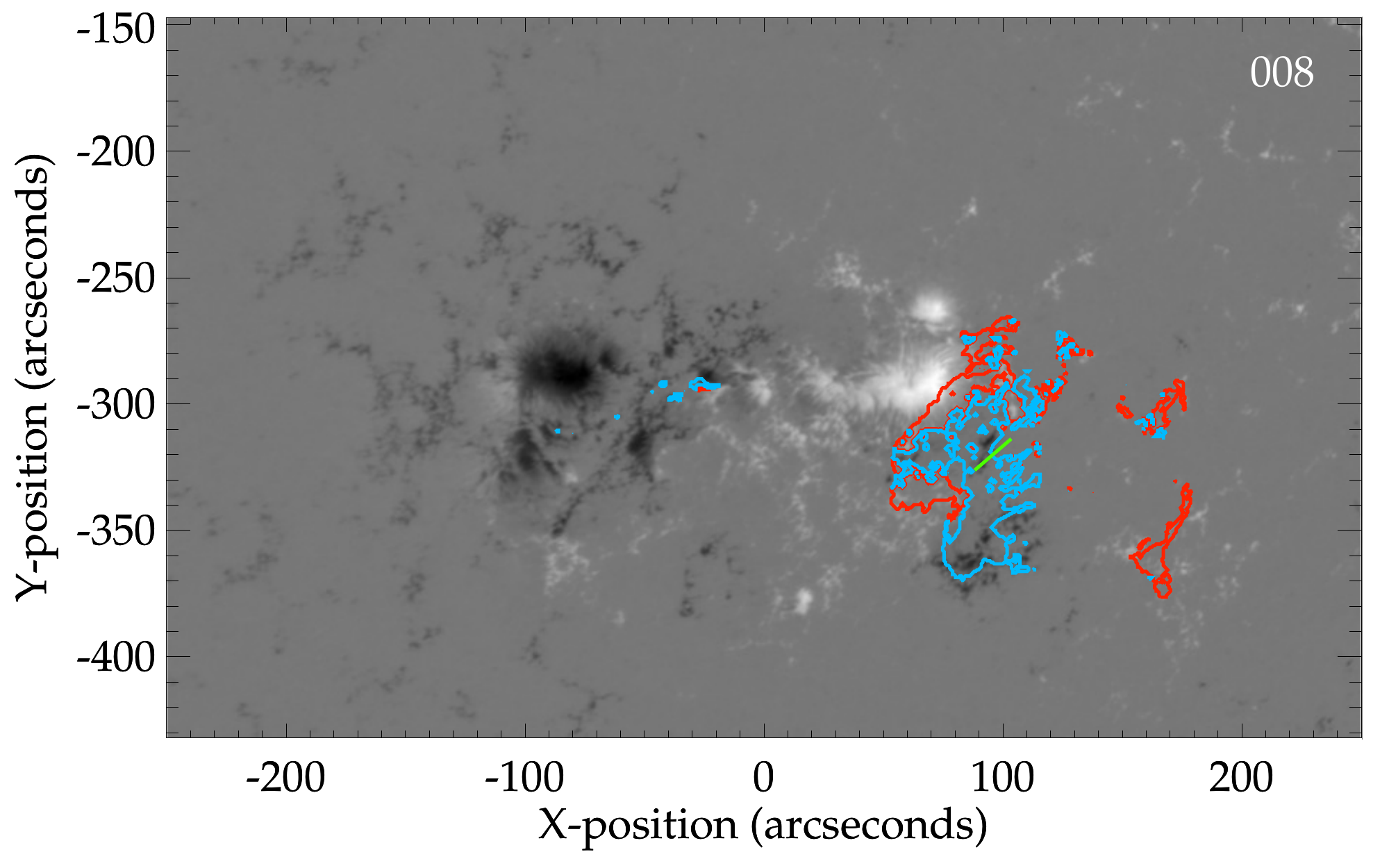}
	\hspace*{0.5cm}
	\includegraphics[width=0.38\linewidth]{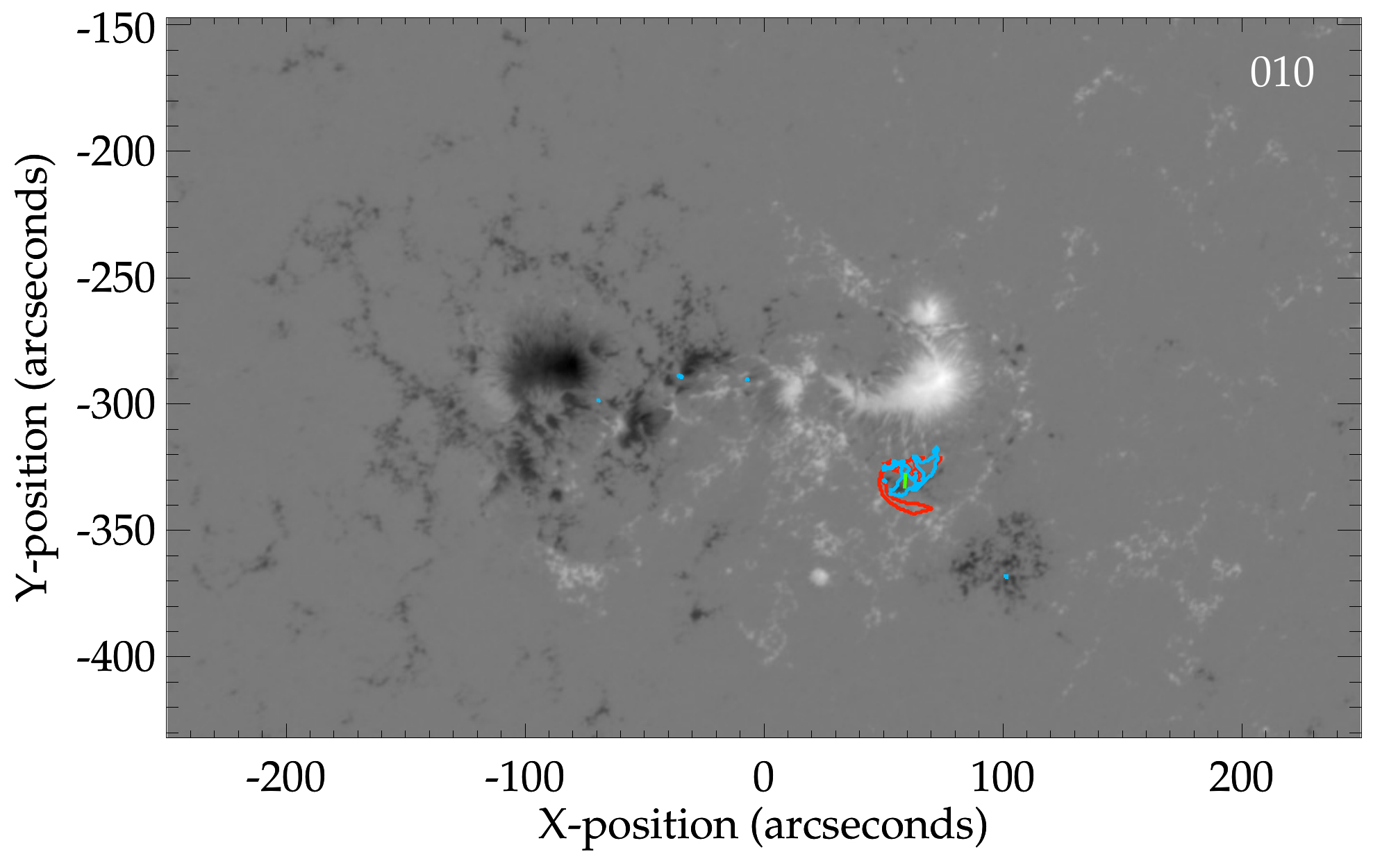} \\
	\includegraphics[width=0.38\linewidth]{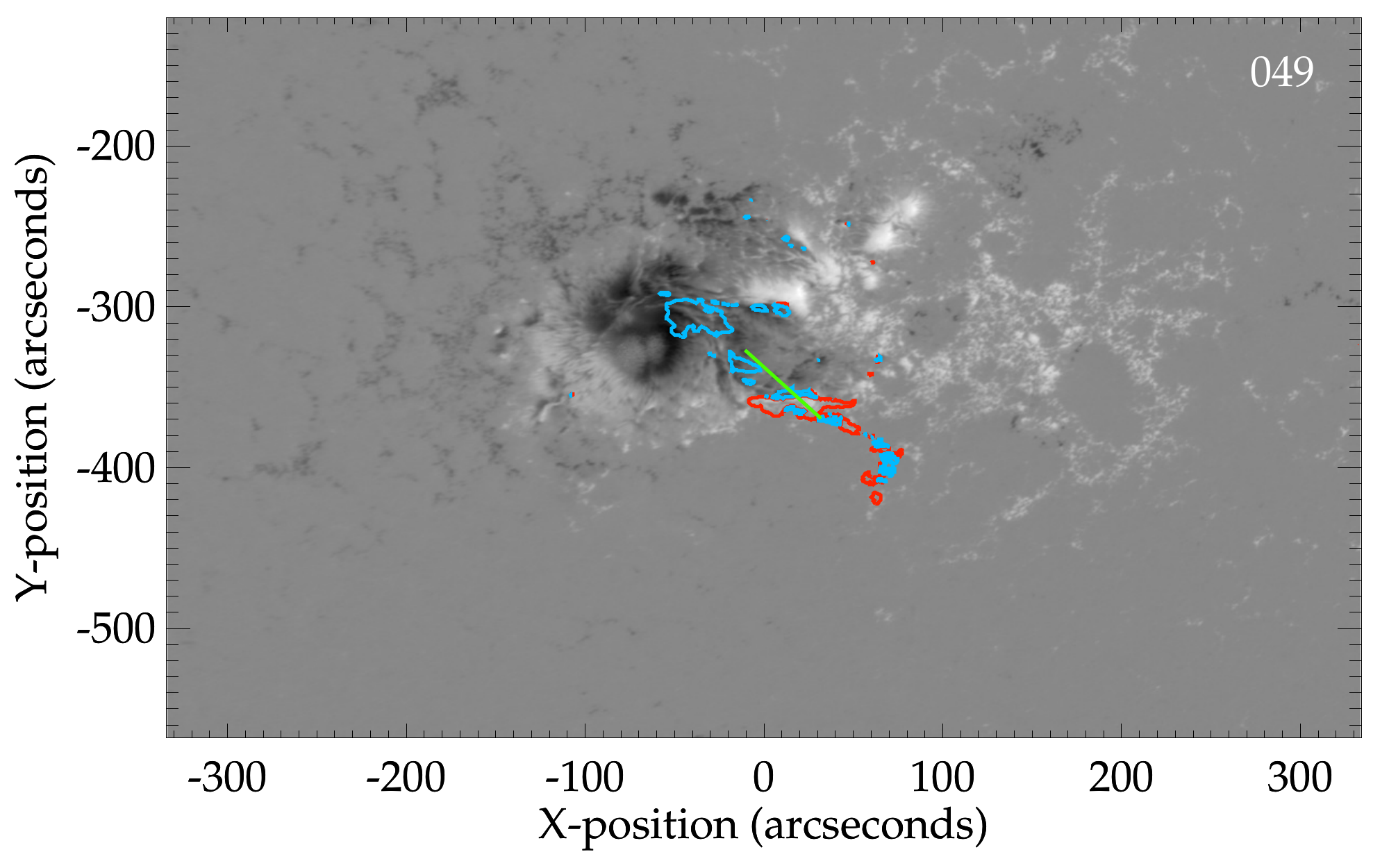}
	\hspace*{0.5cm}
	\includegraphics[width=0.38\linewidth]{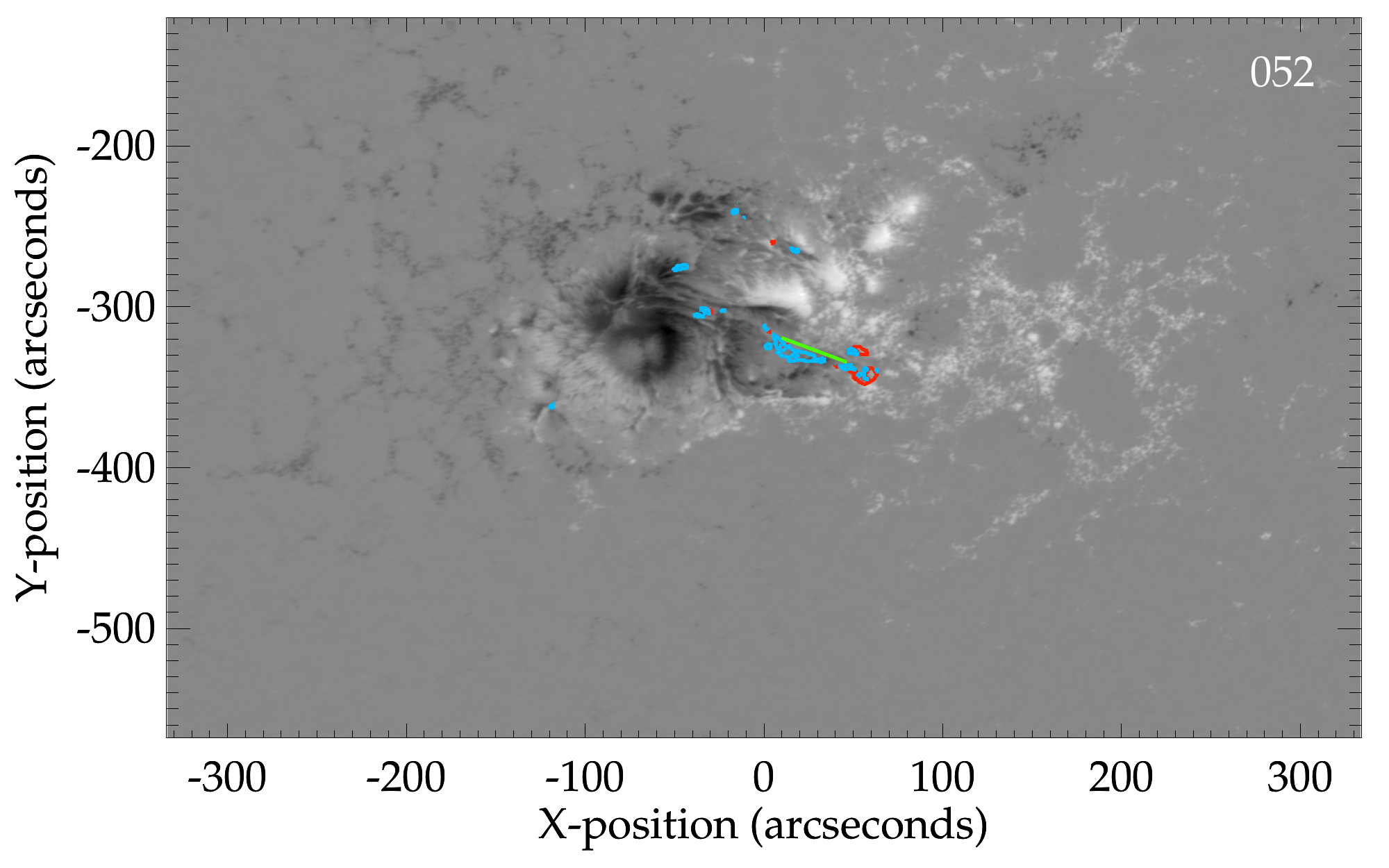} \\
	\includegraphics[width=0.38\linewidth]{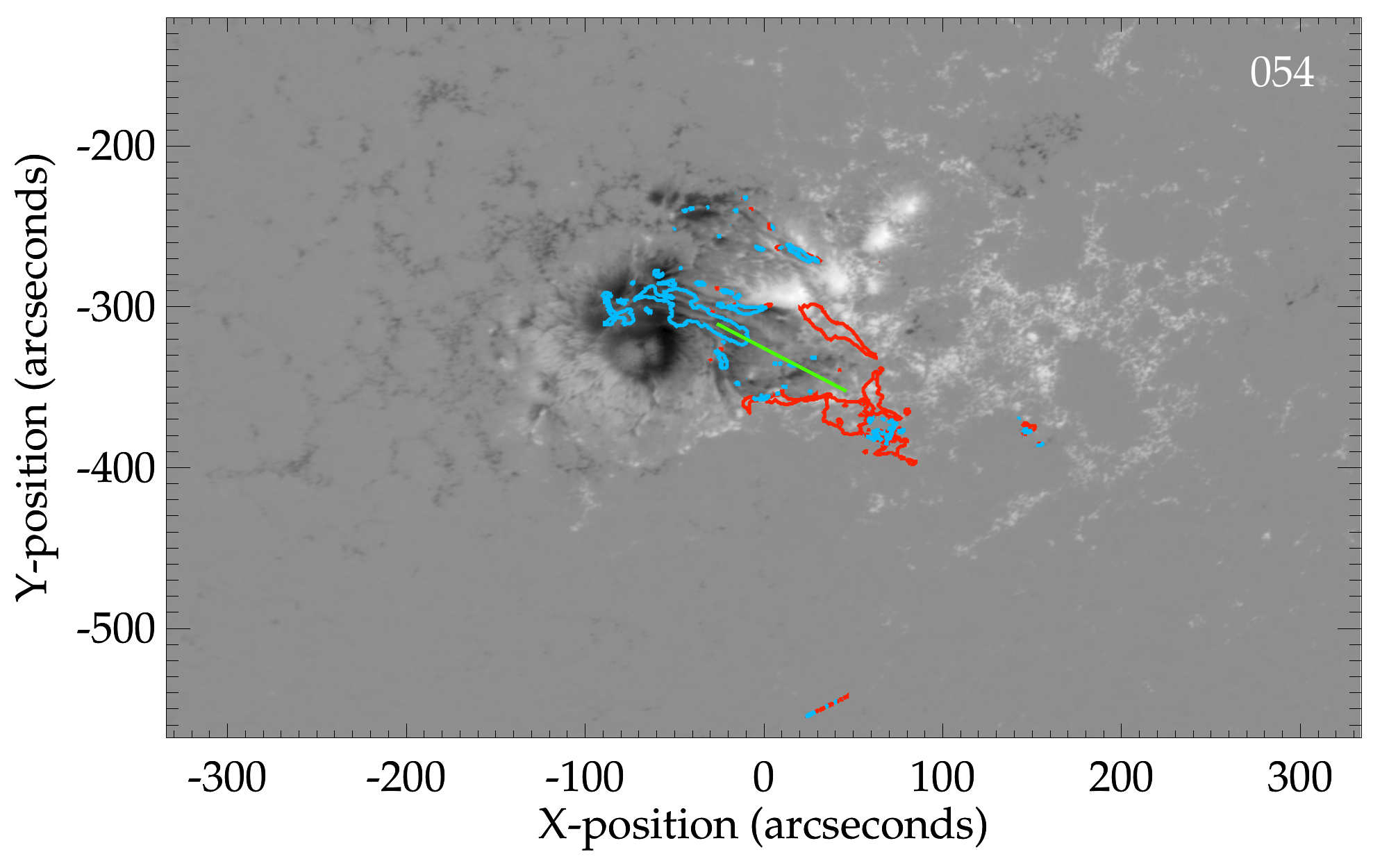}
	\hspace*{0.5cm}
	\includegraphics[width=0.38\linewidth]{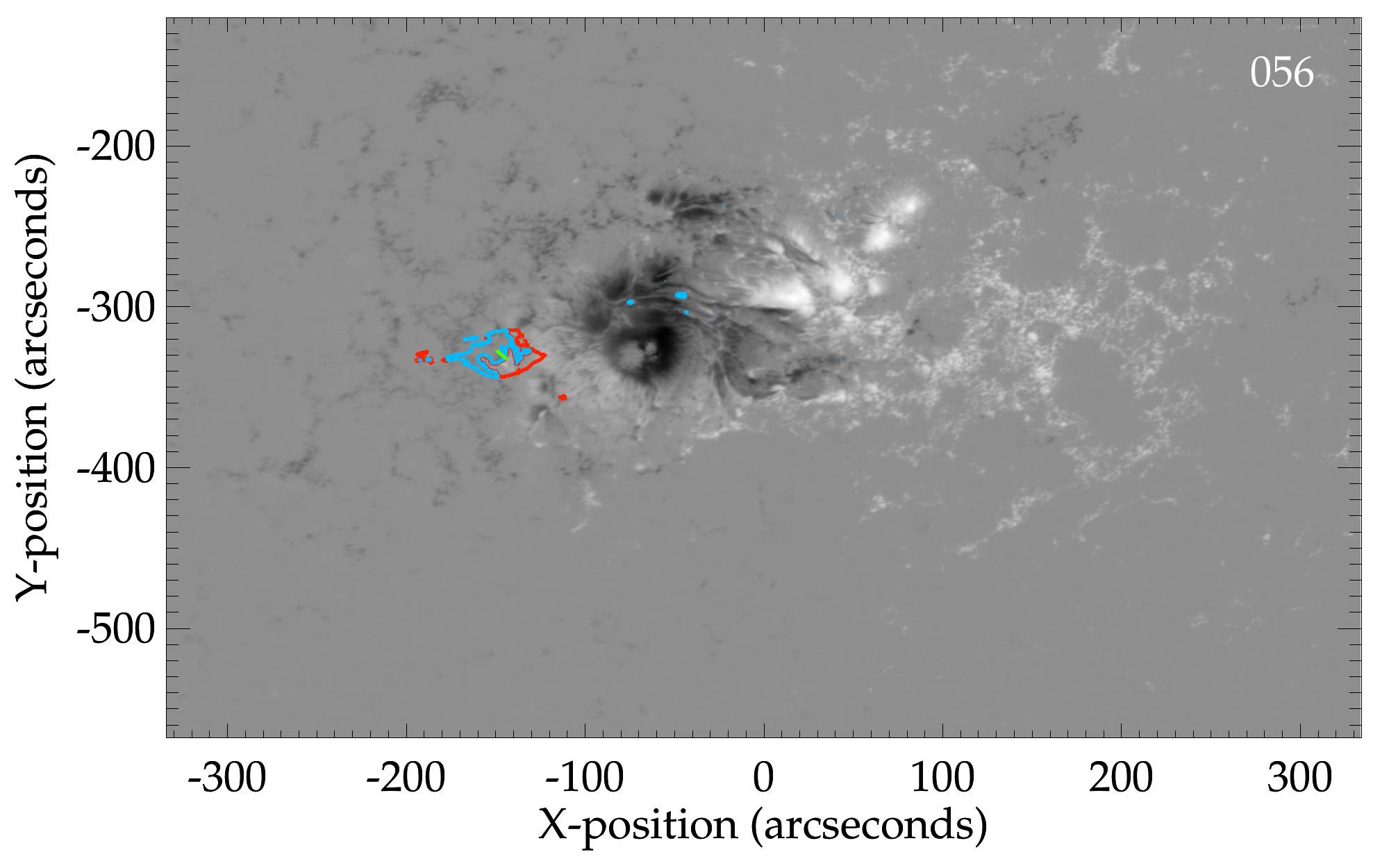} \\
	\includegraphics[width=0.38\linewidth]{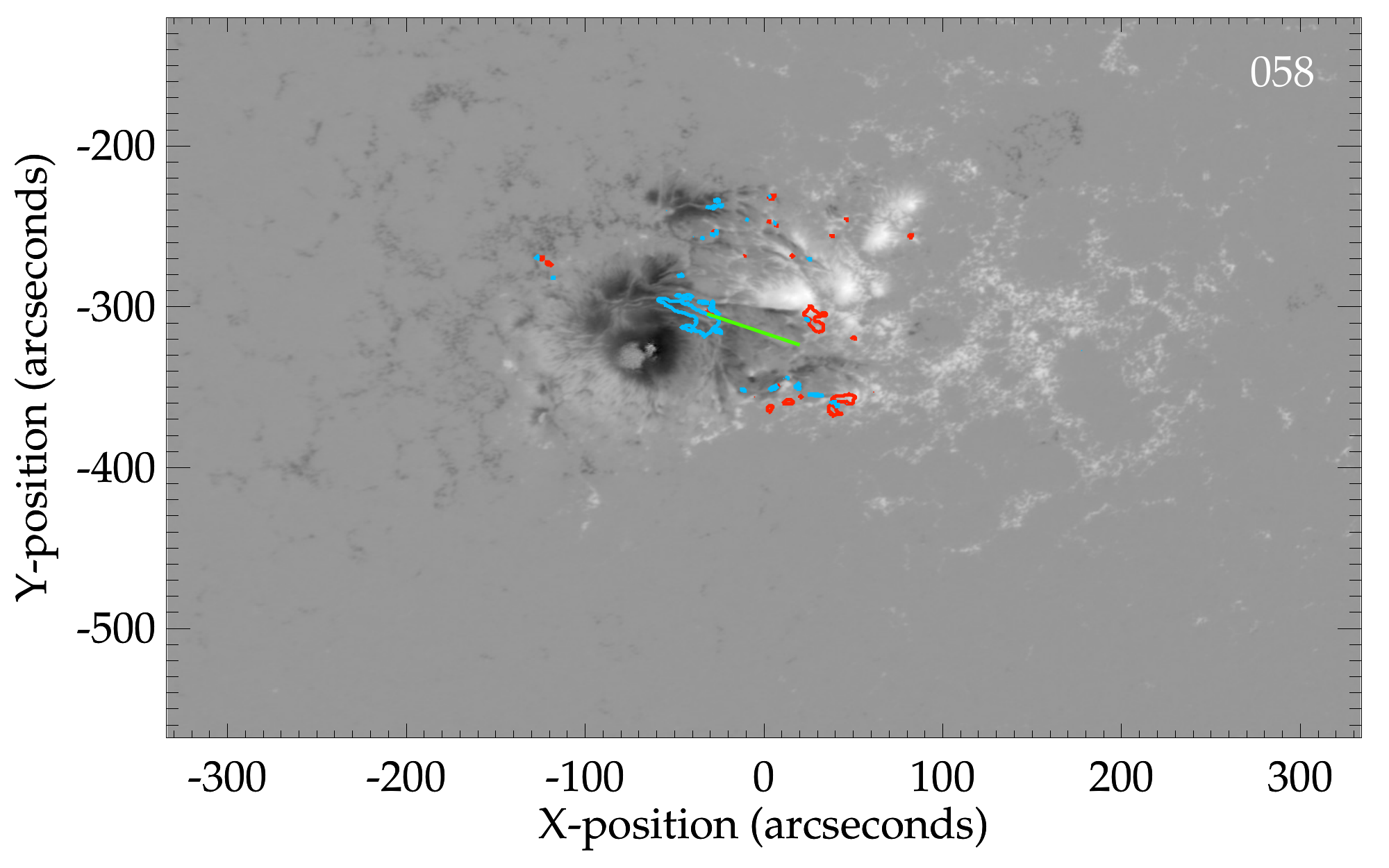}
	\hspace*{0.5cm}
	\includegraphics[width=0.38\linewidth]{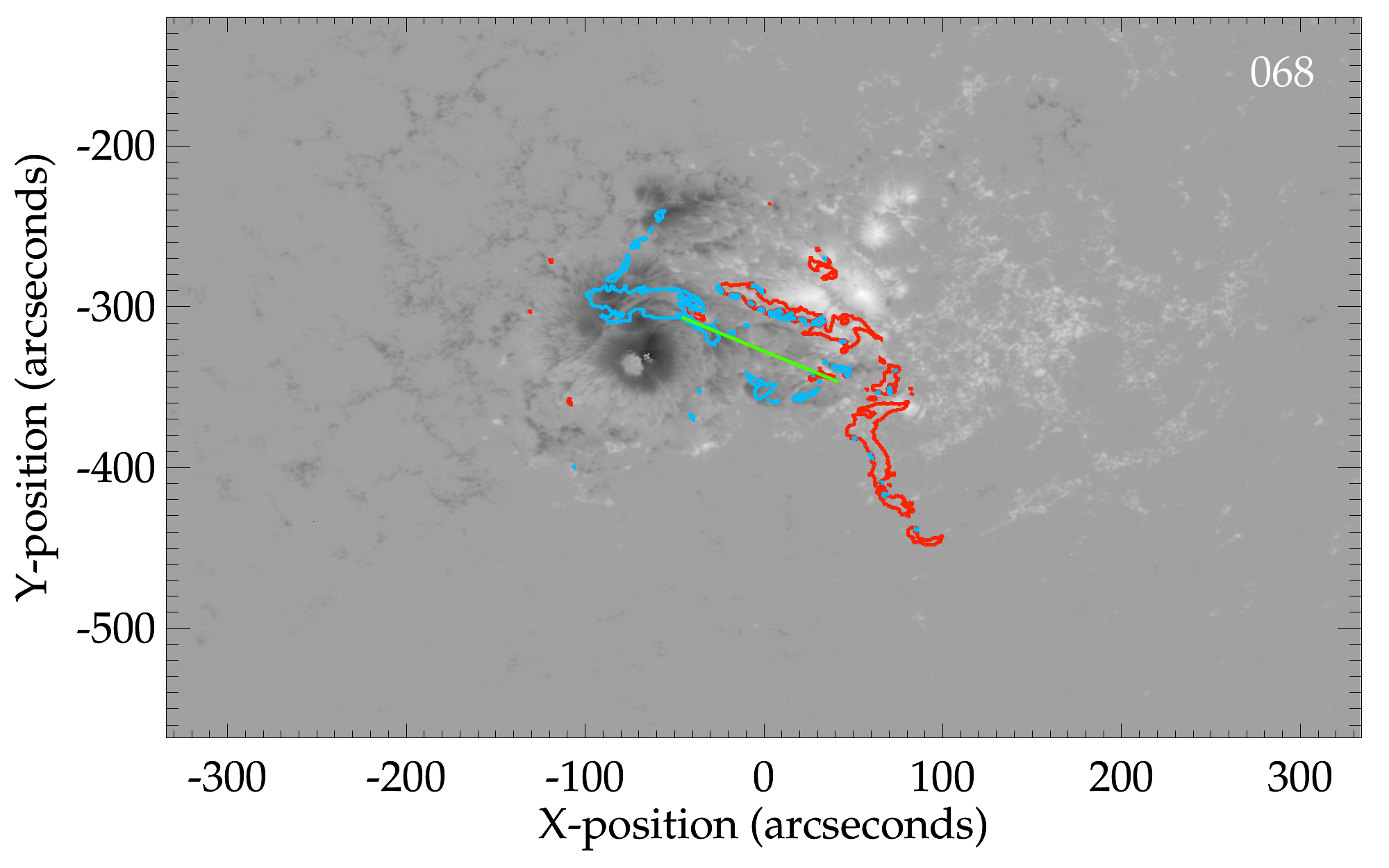} \\
	\includegraphics[width=0.38\linewidth]{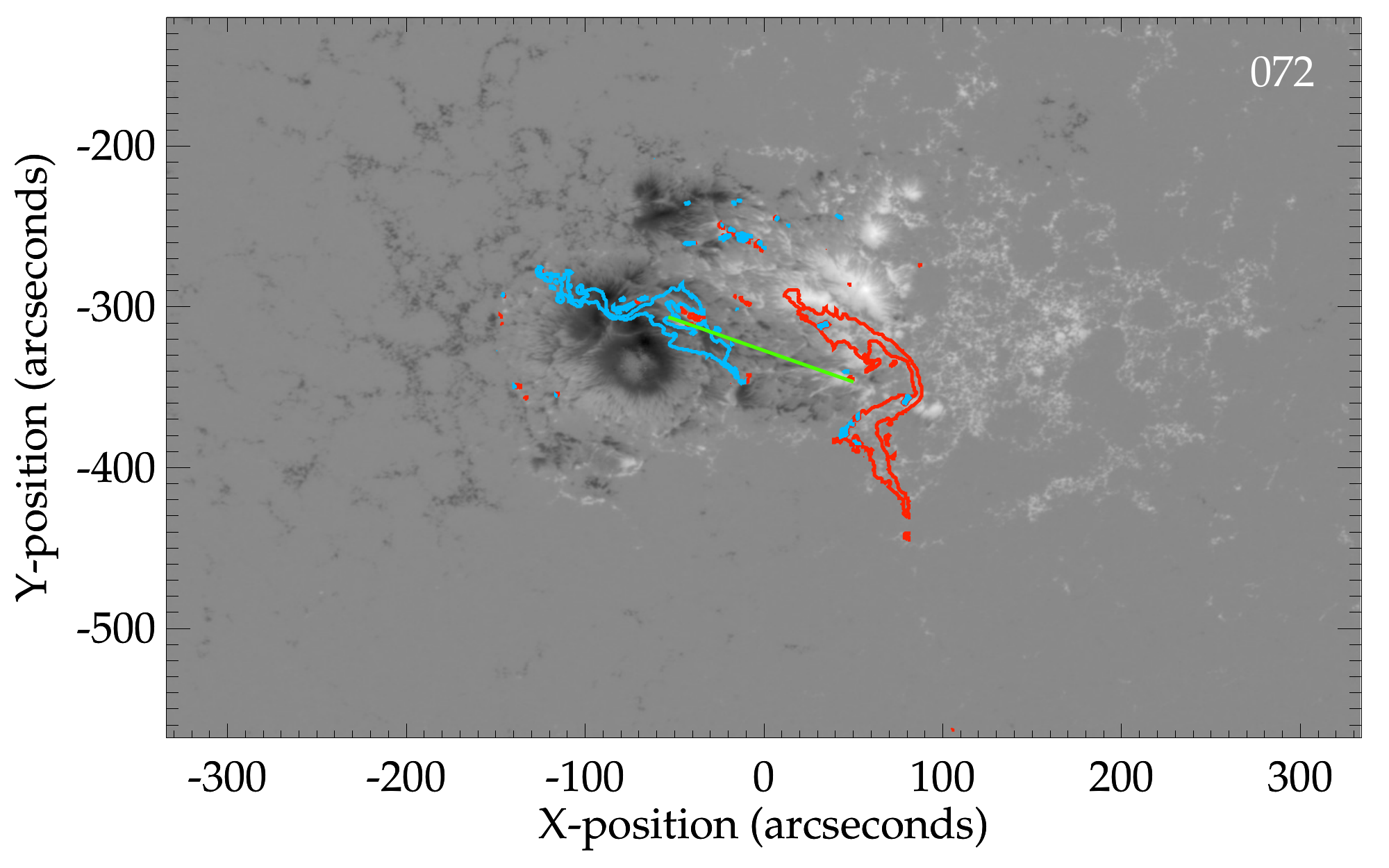}
	\hspace*{0.5cm}
	\includegraphics[width=0.38\linewidth]{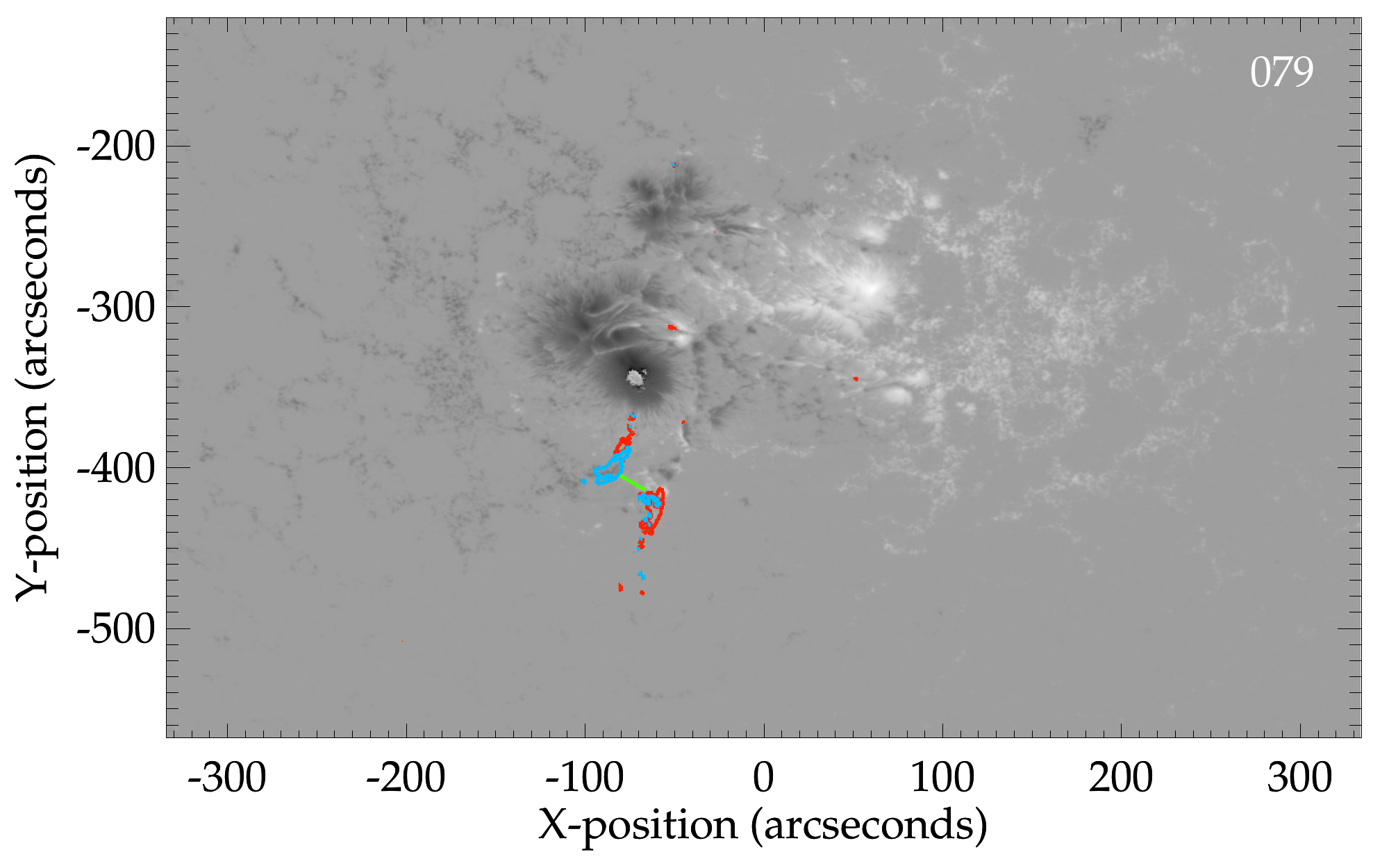}
	\caption{HMI magnetograms for all flares listed in Table~\ref{tab:qppflares}, showing the host active region at the time of the flare peak in the GOES 1--8\,\AA\ waveband. The red and blue contours show the positions of the composite flare ribbons with positive and negative magnetic polarity, respectively. The green lines join the geometric centroids of the positive and negative polarity ribbon components.}
	\label{fig:appendix}
\end{figure}

\begin{figure}\ContinuedFloat
	\centering
	\includegraphics[width=0.38\linewidth]{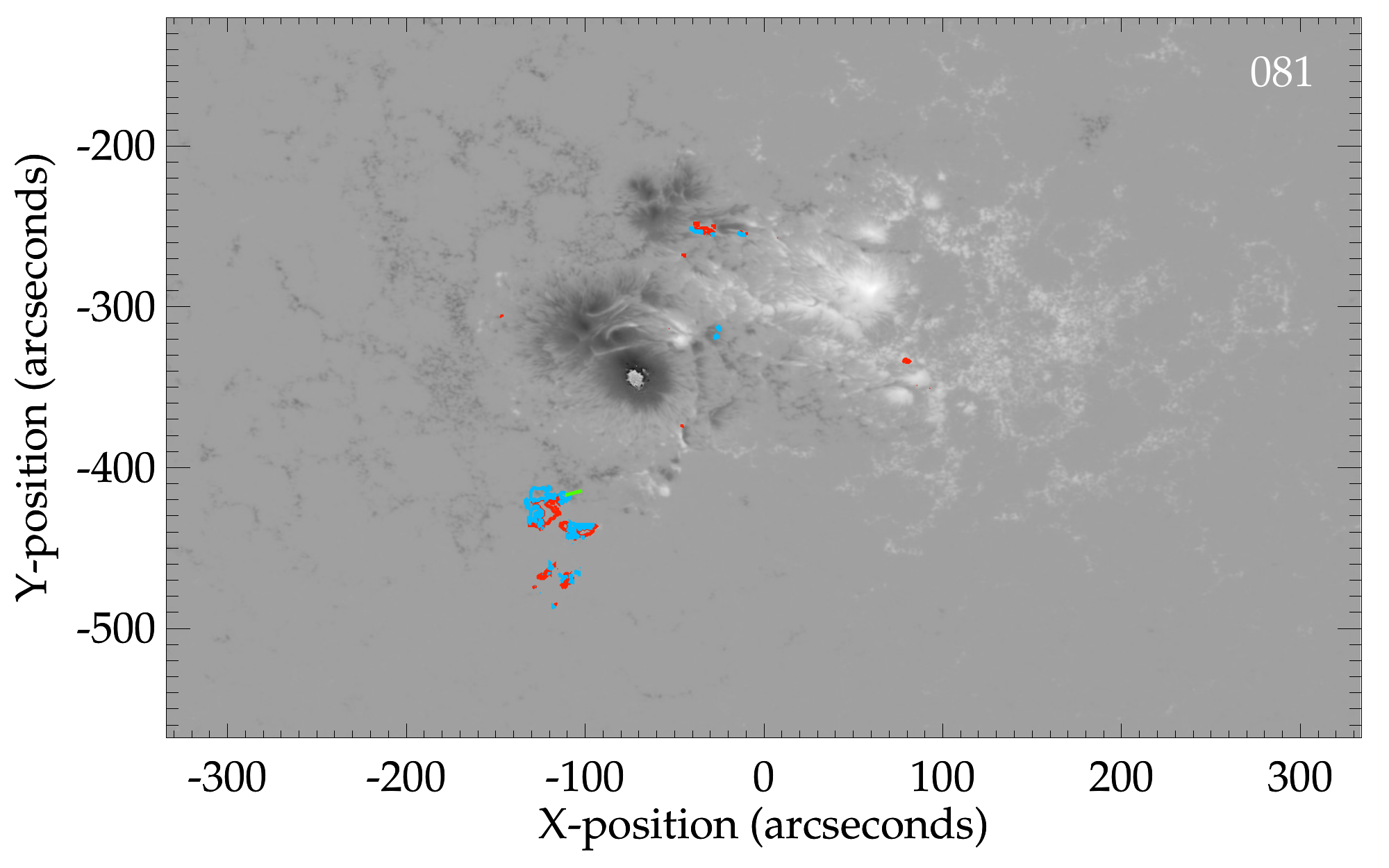}
	\hspace*{0.5cm}
	\includegraphics[width=0.38\linewidth]{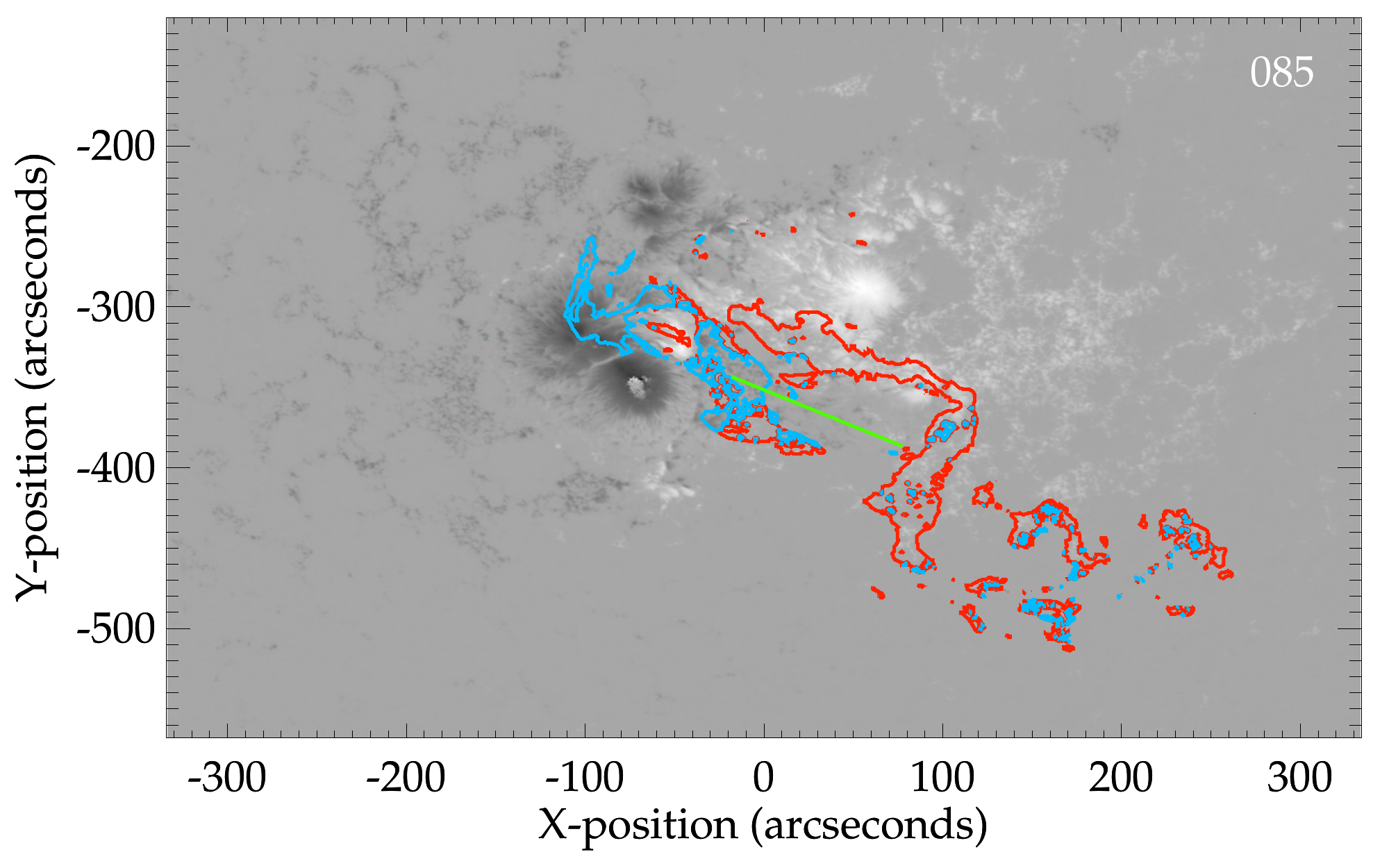} \\
	\includegraphics[width=0.38\linewidth]{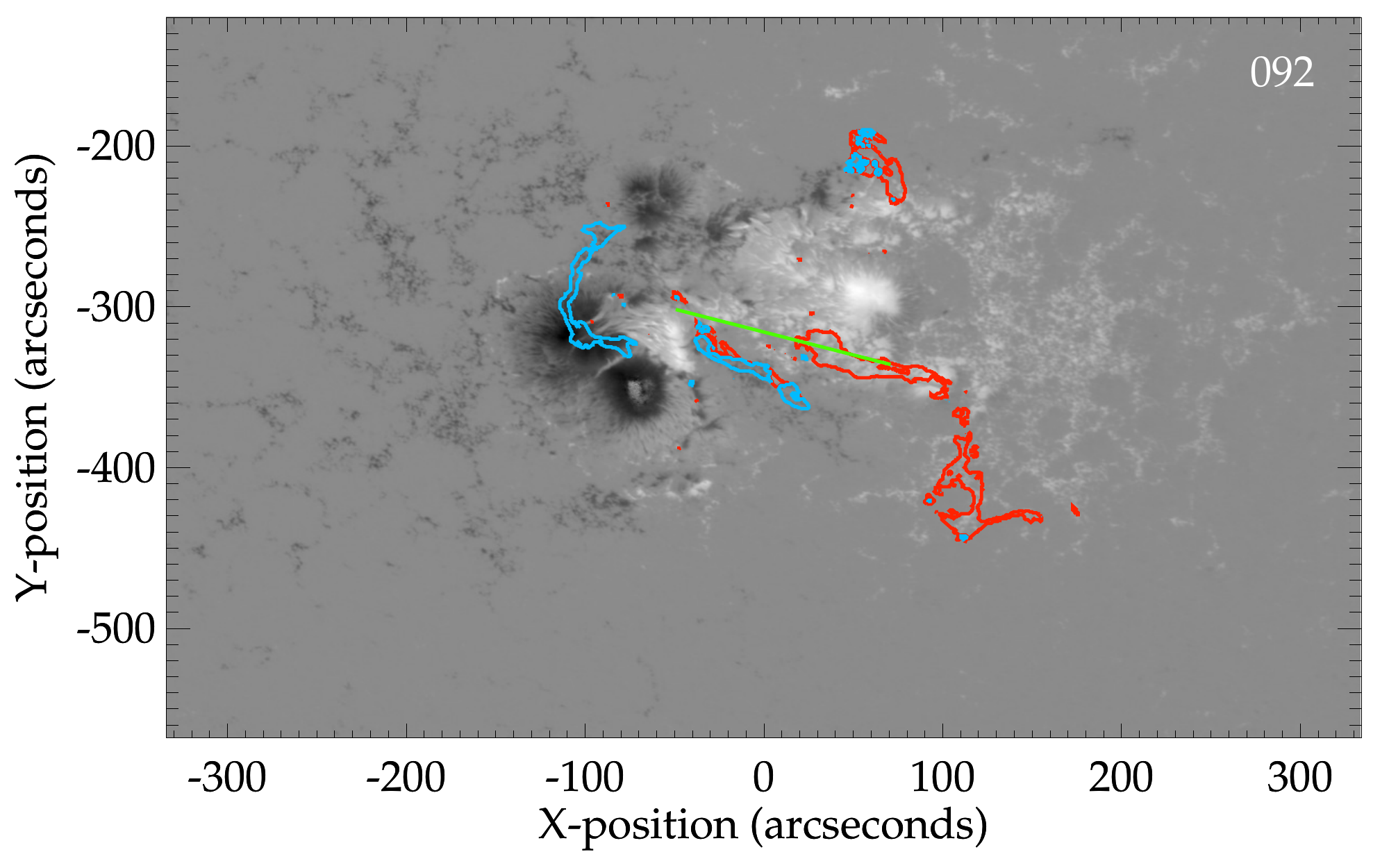}
	\hspace*{0.5cm}
	\includegraphics[width=0.38\linewidth]{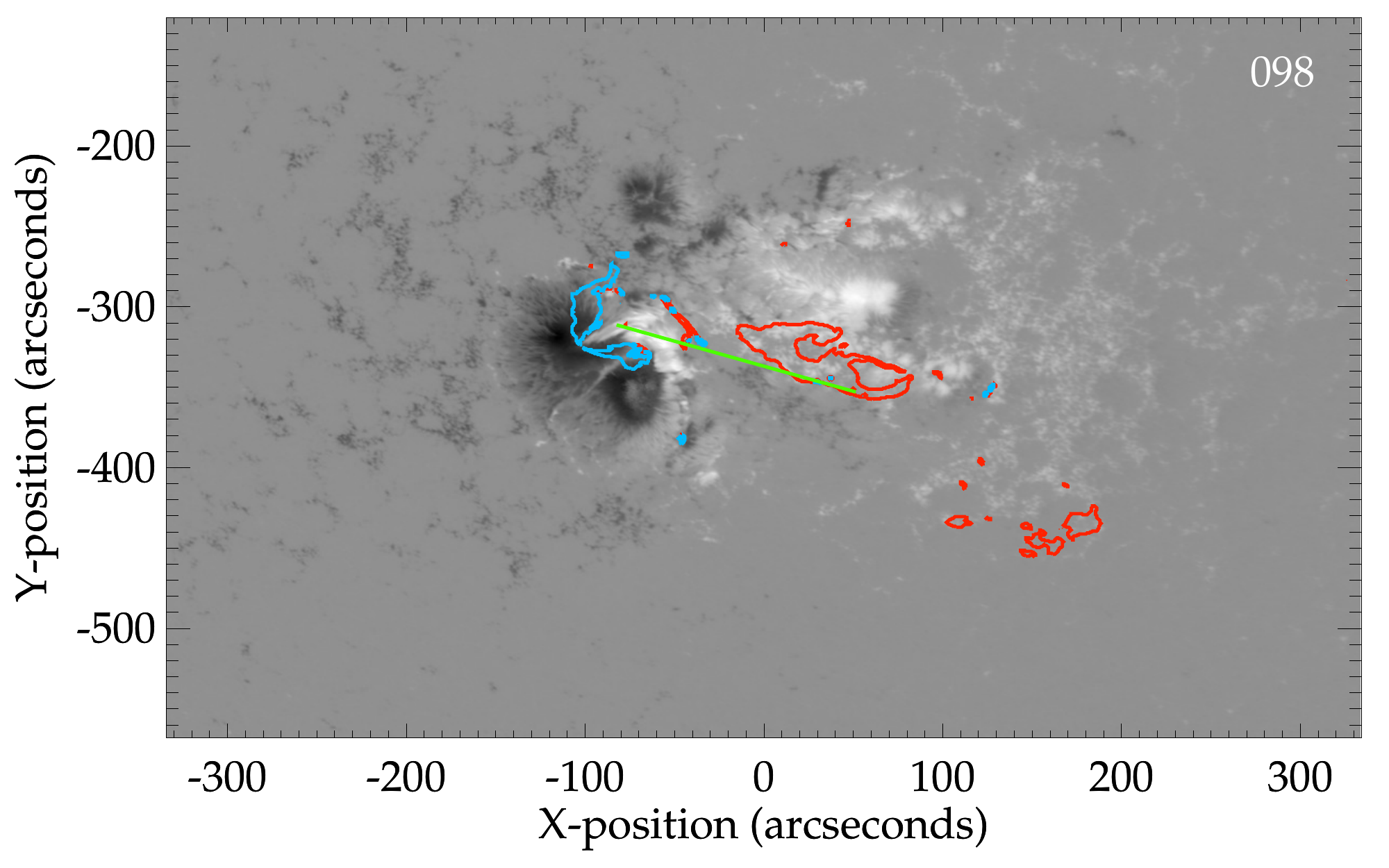} \\
	\includegraphics[width=0.38\linewidth]{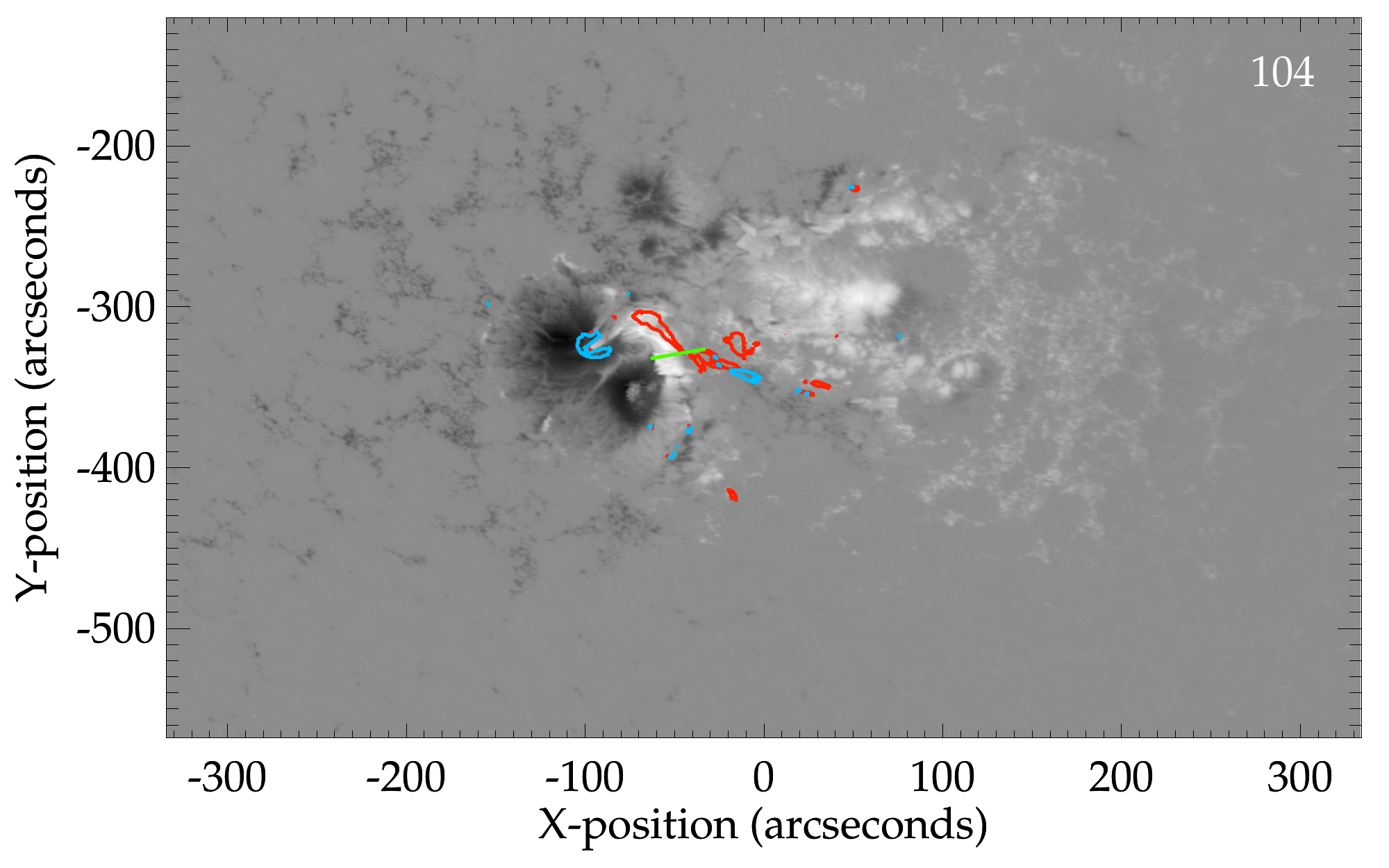}
	\hspace*{0.5cm}
	\includegraphics[width=0.38\linewidth]{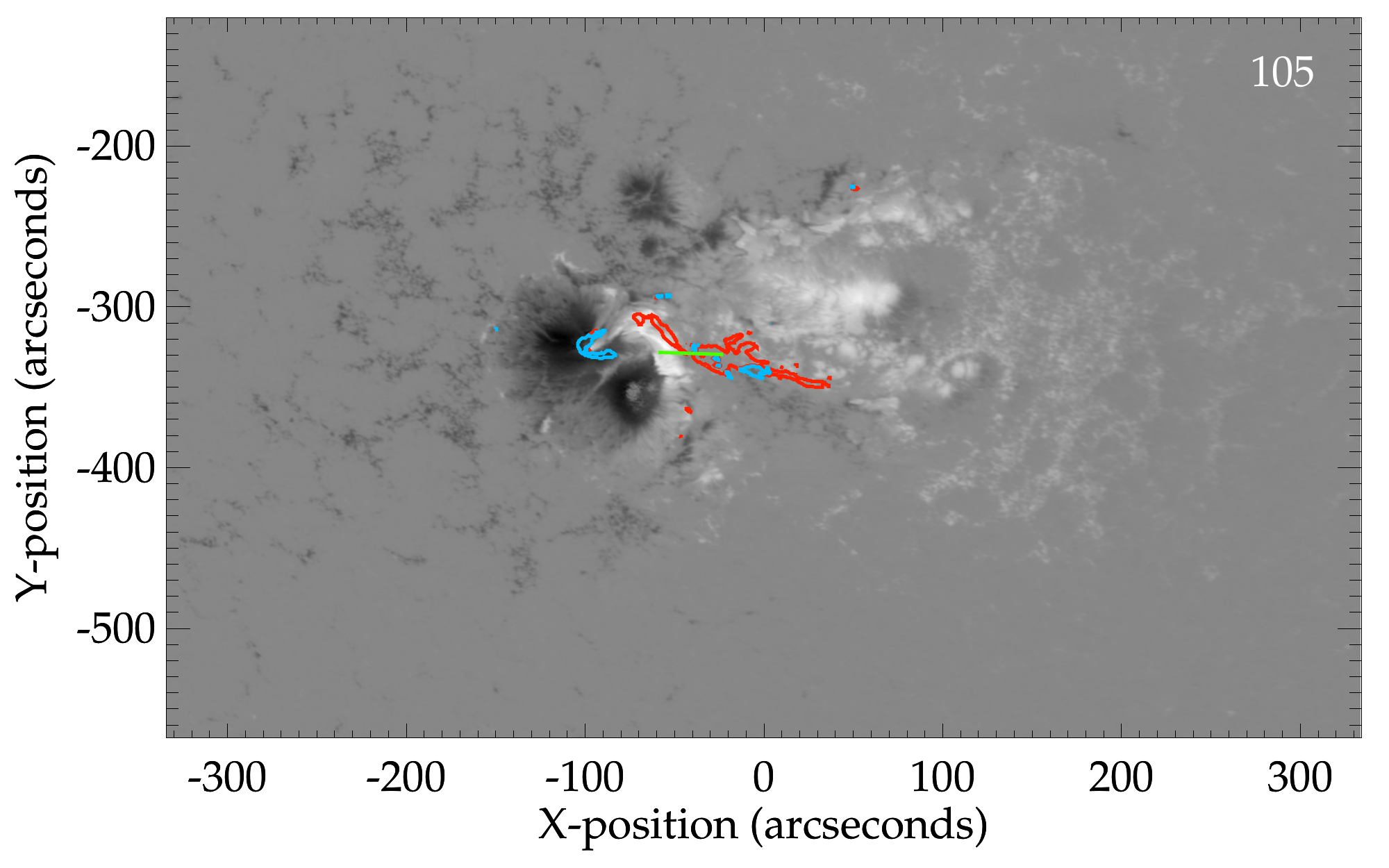} \\
	\includegraphics[width=0.38\linewidth]{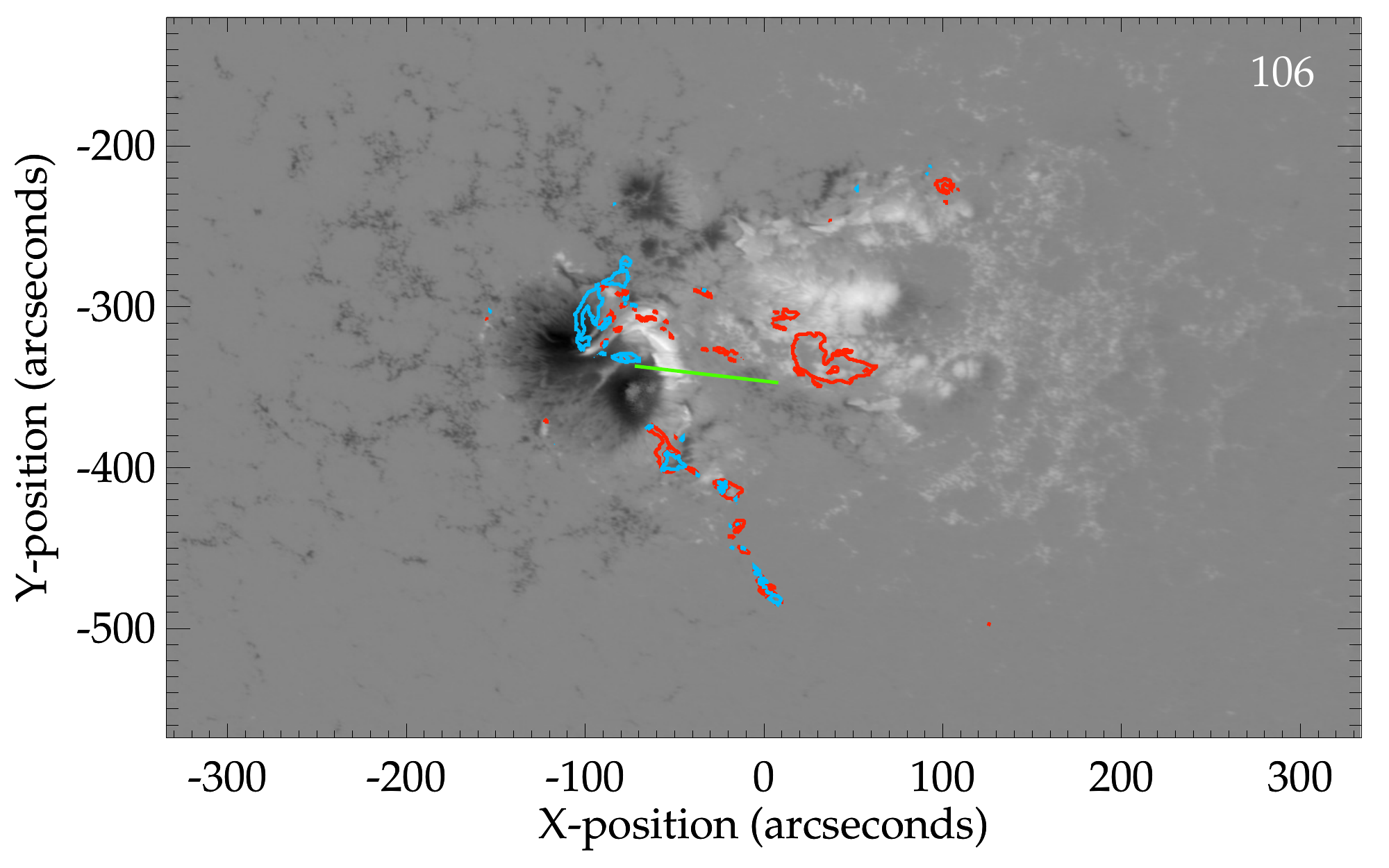}
	\hspace*{0.5cm}
	\includegraphics[width=0.38\linewidth]{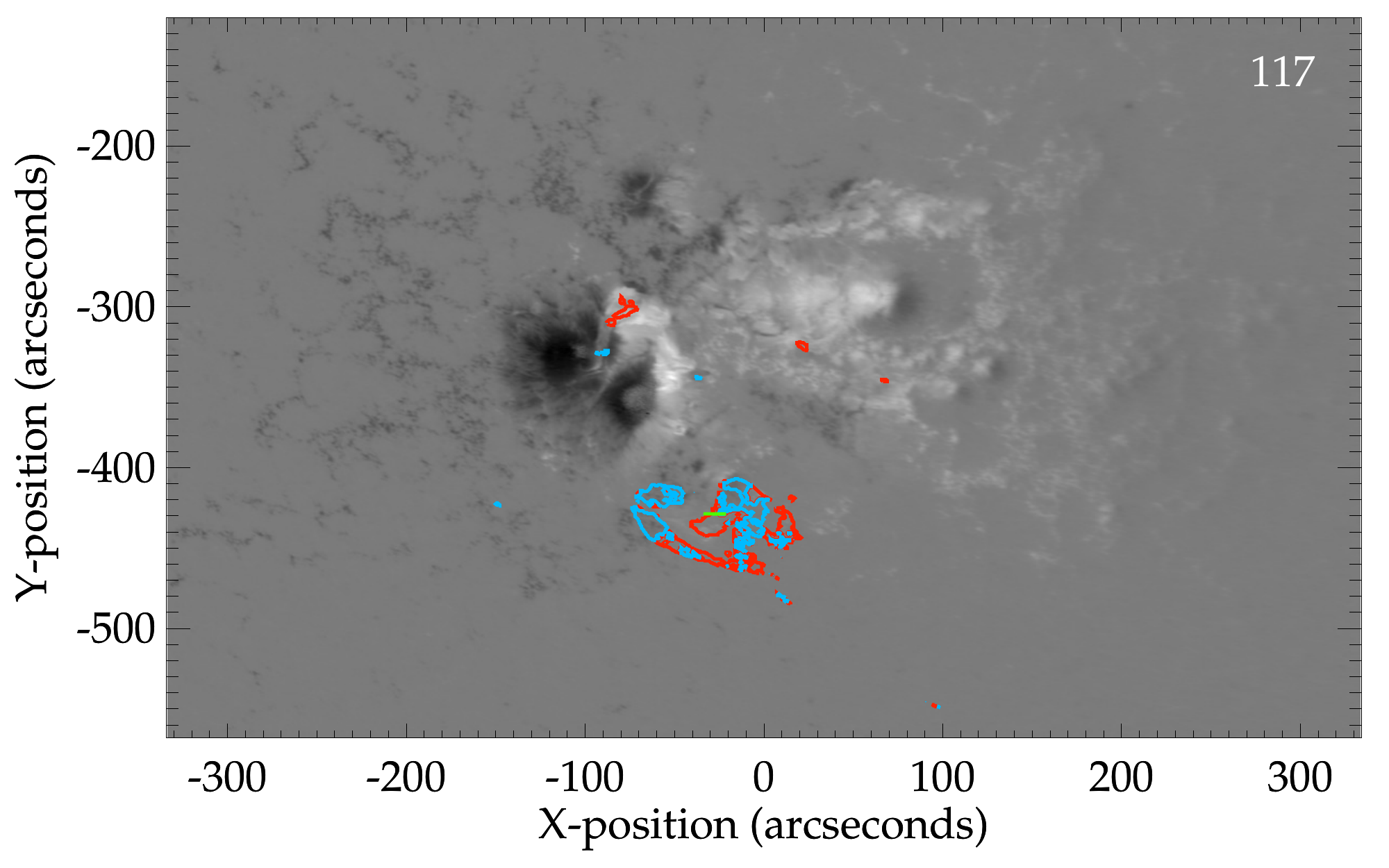} \\
	\includegraphics[width=0.38\linewidth]{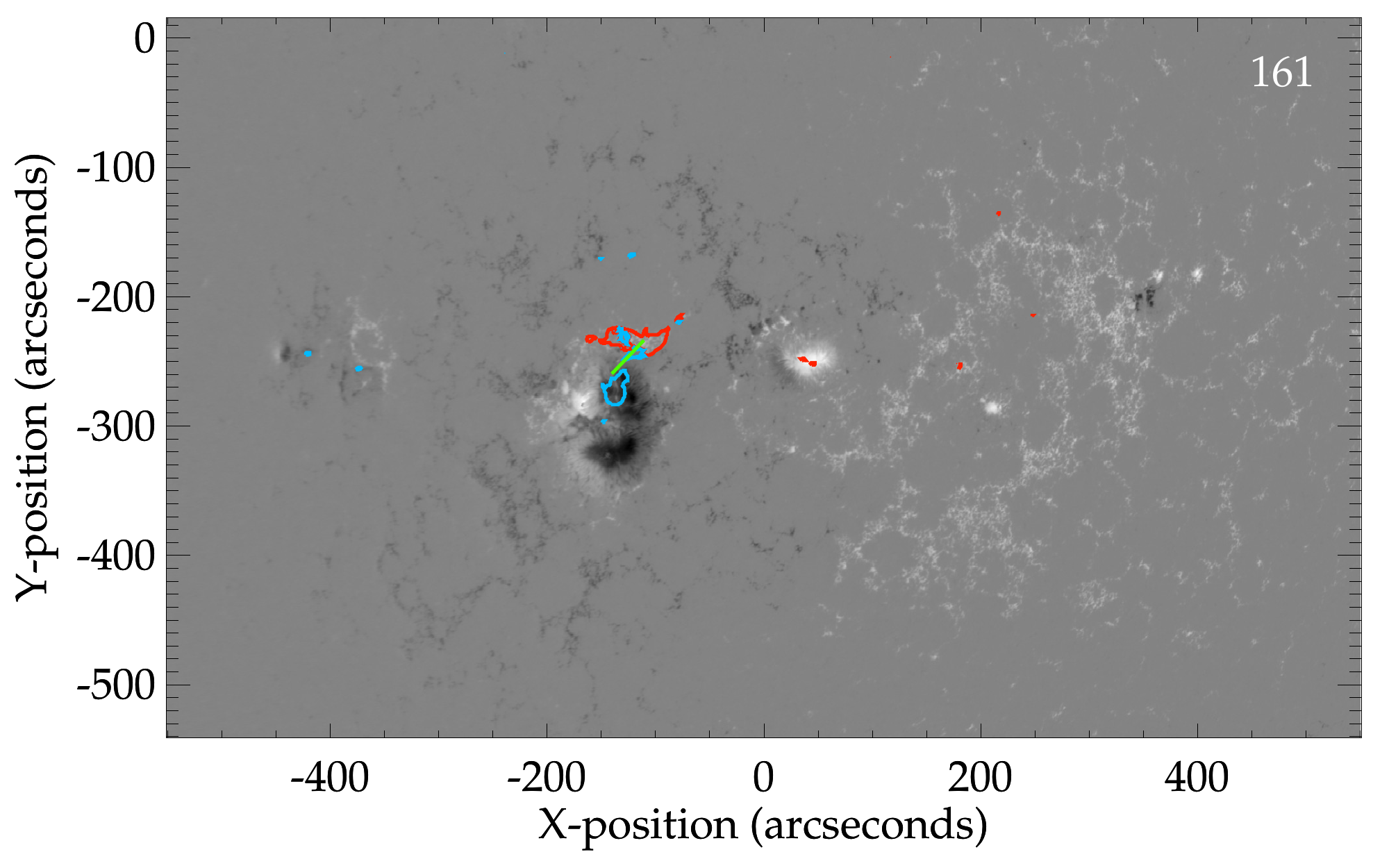}
	\hspace*{0.5cm}
	\includegraphics[width=0.38\linewidth]{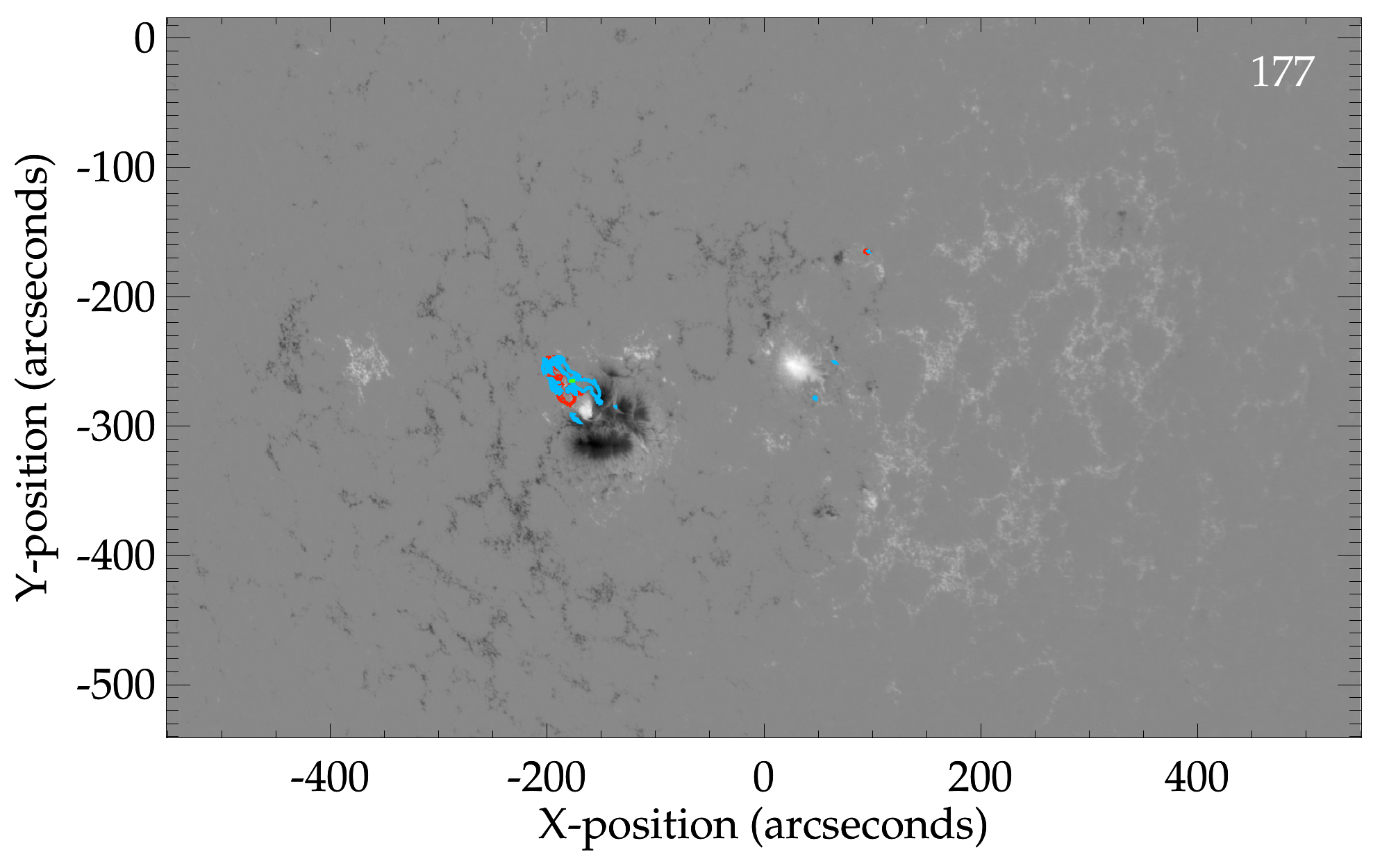}
	\caption{\emph{Continued.}}
\end{figure}

\clearpage

\section{Additional tables}
\label{sec:tab}

\begin{table}[!h]
\centering
\caption{Summary of the flare properties. The first column contains a numerical label for the flares, the second column is the duration of the flare impulsive phase, the third column is the area of the flare ribbons, the fourth is the separation distance of the ribbons, the fifth is the average magnetic field strength measured at the photosphere below the flare ribbons, and the sixth is the total unsigned magnetic flux below the ribbons.}
\label{tab:appendix}
\begin{tabular}{c c c c c c}
	\hline\hline
	\rule{0pt}{10pt} Flare no. & Flare duration (s) & $S_{\mathrm{ribbon}}$ (Mm$^2$)	& $d_{\mathrm{ribbon}}$ (Mm) & $B_{\mathrm{ribbon}}$ (G) & $|\Phi|_{\mathrm{ribbon}}$ (10$^{20}$\,Mx) \\[2.3pt]
	\hline
	\rule{0pt}{10pt} 008 & 714 	& 2162.1 	& 14.4 	& 131.7 & 28.5	\\[2.3pt] 
	010 & 131 	& 173.0 	& 4.7 	& 136.9 & 2.4 	\\[2.3pt]
	049 & 650	& 1042.2 	& 44.8 	& 396.2 & 41.3	\\[2.3pt]
	052 & 158 	& 245.6 	& 28.7 	& 216.1 & 5.3 	\\[2.3pt]
	054 & 2214 	& 1531.3 	& 61.2 	& 437.5 & 67.0 	\\[2.3pt]
	056 & 406 	& 522.0 	& 5.8 	& 114.2 & 6.0 	\\[2.3pt]
	058 & 658 	& 411.7 	& 40.0 	& 515.0 & 21.2	\\[2.3pt]
	068 & 2613 	& 1977.6 	& 70.0 	& 522.0 & 103.2	\\[2.3pt]
	072 & 1575 	& 1945.3 	& 80.5 	& 486.5 & 94.6	\\[2.3pt]
	079 & 537 	& 289.3 	& 12.4 	& 150.5 & 4.4 	\\[2.3pt]
	081 & 248 	& 373.3 	& 7.2 	& 80.7	& 3.0 	\\[2.3pt]
	085 & 3270 	& 6367.6	& 76.8 	& 272.8 & 174.7	\\[2.3pt]
	092 & 2000 	& 2249.8 	& 90.3 	& 373.5 & 84.0	\\[2.3pt]
	098 & 1300 	& 1880.8 	& 101.2	& 420.2 & 79.0  \\[2.3pt]
	104 & 449 	& 425.8 	& 22.6 	& 541.5 & 23.1	\\[2.3pt]
	105 & 510 	& 589.2 	& 26.1 	& 394.4 & 23.2 	\\[2.3pt]
	106 & 1664 	& 1209.5 	& 58.2 	& 346.3 & 41.9	\\[2.3pt]
	117 & 638 	& 1291.0 	& 9.0 	& 82.2	& 10.6 	\\[2.3pt]
	161 & 803 	& 736.7 	& 29.1 	& 358.7 & 26.4	\\[2.3pt]
	177 & 333 	& 515.4 	& 4.1 	& 169.2 & 8.7	\\[2.3pt]
	\hline
\end{tabular}
\end{table}

\end{appendix}

\end{document}